\documentclass[pra,amssymb,amsmath,showpacs]{revtex4}

\usepackage{graphicx}
\usepackage{amssymb}
\usepackage{amsmath}
\usepackage{color}

\usepackage[varg]{txfonts}	

\def\url#1{\textcolor{blue}{\underline{#1}}}	
\definecolor{oneblue}{rgb}{0,0,0.65}
\definecolor{onered}{rgb}{0.65,0,0}

\def\url#1{\textcolor{blue}{\underline{#1}}}	
\def\corr#1{\textcolor{black}{#1}}    		

\def\b#1{\textcolor{oneblue}{#1}} 

 \large\normalsize	 
\begin{document}

\def\tit{Generalized cable theory for neurons in complex and
  heterogeneous media}

\title{\tit}

\author{Claude Bedard and Alain Destexhe \\ \ \\ Unit\'e de
Neuroscience, Information et Complexit\'e (UNIC), \\ CNRS,
Gif-sur-Yvette, France}

\date{\today}




\begin{abstract}

  Cable theory has been developed over the last decades, usually
  assuming that the extracellular space around membranes is a perfect
  resistor.  However, extracellular media may display more complex
  electrical properties due to various phenomena, such as
  polarization, ionic diffusion or capacitive effects, but their
  impact on cable properties is not known.  In this paper, we
  generalize cable theory for membranes embedded in arbitrarily
  complex extracellular media.  We outline the generalized cable
  equations, then consider specific cases.  The simplest case is a
  resistive medium, in which case the equations recover the
  traditional cable equations.  We show that for more complex media,
  for example in the presence of ionic diffusion, the impact on cable
  properties such as voltage attenuation can be significant.  We
  illustrate this numerically always by comparing the generalized
  cable to the traditional cable.  We conclude that the nature of
  intracellular and extracellular media may have a strong influence on
  cable filtering as well as on the passive integrative properties of
  neurons.

\end{abstract}

\maketitle


\section{Introduction}

Cable theory, initially developed by Rall \cite{Rall1962}, is one of
the most significant contributions of theoretical neuroscience and has
been extremely useful to explain a large range of phenomena (reviewed
in \cite{Rall1995}).  However, cable theory makes a number of
assumptions, one of which is that the extracellular space around
neurons can be modeled by a resistance, or in other words, that the
medium around neurons is resistive or ohmic.  While some measurements
seem to confirm this assumption~\cite{Logothetis}, other measurements
revealed a marked frequency dependence of the extracellular
resistivity~\cite{Gabriel1996a,Gabriel1996b}, which indicates that the
medium is non-resistive.  Indirect measurements of the extracellular
impedance also show evidence for deviations from
resistivity~\cite{BedDes2006a,BedDes2010,Deg2010,Baz2011}, which could be
explained by the influence of ionic diffusion~\cite{BedDes2011a}.
Despite such evidence for non-resistive media, the possible impact on
cable properties has not been evaluated.

The effect of non-resistive media can be investigated by integrating
this effect in the impedance of the extracellular medium, $Z_e$, and
in particular, through its frequency dependence.  For example, it can
be shown that $Z_e \sim 1/\omega$ for capacitive effects or electric
polarization~\cite{BedDes2006b}, $Z_e \sim 1/\sqrt{\omega}$ for ionic
diffusion (also called the ``Warburg impedance''~\cite{BedDes2009}),
while $Z_e$ would be constant for a perfectly resistive medium.  To
integrate such effects in a given formalism, such as the genesis of
extracellular potentials, our approach has been to integrate a general
frequency-dependent function $Z_e(\omega)$ in the formalism, and then
consider specific cases~\cite{BedDes2009,BedDes2011a}.

In the present paper, we follow this approach and generalize cable
equations for media with arbitrarily complex frequency-dependent
impedance.  With numerical simulations, we consider specific cases
such as resistive media, ionic diffusion, capacitive media, etc.  We
evaluate a number of possible consequences on the variation of the
membrane potential along the cable, and how such effects could be
measured experimentally.


\section{Methods}

\begin{figure}[h!] 
\centering
\includegraphics[width=8cm]{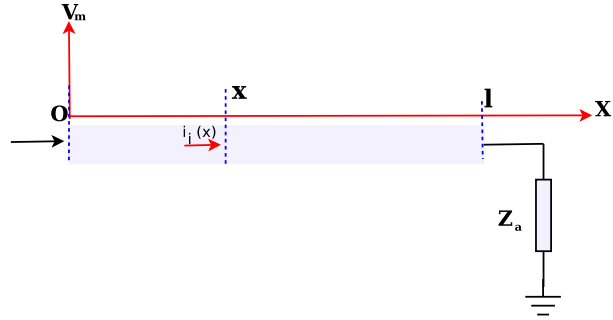}

\caption{\small \corr{(Color online)} Convention used to calculate the
  input impedance and transfer function.  A cable segment of length
  \b{$l$} is represented, with an impedance \b{$Z_a$} in series, at
  the end of the cable.  This ``auxiliary impedance'' \b{$Z_a$} takes
  into account the influence of the other compartments on the axial
  current \b{$i_i$} and transmembrane potential \b{$V_m$} in a compact
  form.  \b{$Z_a=\frac{V_m(l)}{i_i(l)}$} where \b{$i_i(l)$} is the
  current per unit length and \b{$V_m(l)$} is the transmembrane
  voltage at coordinate \b{$x=l$}.  }

\label{fig0}
\end{figure}

All simulations were done using MATLAB.  To simulate the cable
structure of the models, a classic compartmental model strategy was
used for simulations (see Fig.~\ref{fig2}F), but was different from
the one used in common simulator programs such as NEURON (Hines and
Carnevale, \cite{Hines97}).  Each cylindric compartment is connected
to intracellular and extracellular resistances or impedances, and
these are normally used to solve the cable equations. In the present
paper, we used another, equivalent method which consists of defining
an {\it auxiliary impedance}, given by \b{$Z_a=\frac{V_m}{i_i}$} where
\b{$V_m$} and \b{$i_i$} are respectively the transmembrane potential
and the axial current per unit length at the point where \b{$Z_a$} is
connected (see Fig.~\ref{fig0}). This auxiliary impedance allows to
take into account the influence of other compartments, including the
soma, over the axial current and transmembrane potential.  It is
mathematically equivalent to consider the continuity conditions on
axial current and transmembrane potential.

The electric and geometric parameters are considered constant in each
compartment, but are allowed to vary between compartments.  In these
conditions, \b{$V_m$} and \b{$i_i$} are solution of partial
differential equations (cable equations) and thus depend on spatial
coordinates.

The cable equations simulated in this article are generalized to
allow one to include media with complex electrical properties. We
have designed a MATLAB code that simulates such generalized cable
structures, using different types of linear density of complex
impedances (\b{$ [\Omega/m] $}) and specific impedances (\b{$
[\Omega.m] $}) in each compartment. See Results for details of this
method.

All computations were made in Fourier space.  We have applied the
theory to four different types of media to evidence their effect on
the spatial and frequency profile of the membrane potential.  These
models are called SC, FC, FO and NIC, respectively (see
Table~\ref{table1}).  The SC model is the ``standard model'' as
defined by Tuckwell \cite{Tuckwell}; the FC model corresponds to a
model similar to the standard model (based on a closed circuit),
but the cytoplasm and extracellular media impedances can be
frequency-dependent.  The FO type model is the same, with an open
circuit (no return current).  The NIC model includes a non-ideal
capacitance similar to a previous study \cite{BedDes2008}. See
Results for details of these models.

All numerical simulations were made using a ``continuous
ball-and-stick'' model, consisting of a single cylindric compartment,
described as a continuum (see Results), and a spherical soma.  The
dendritic compartment has a radius of \b{$2~\mu m$} and a membrane
time constant of \b{$5~ms$}, which corresponds to typical values of
\text{in vivo} conditions.  The has a radius of \b{$7.5~nm$} and the
specific capacitance was of \b{$0.01~F/m^2$}.  These parameters
represent typical values used in a number of previous studies
\cite{Rall1995,Koch,Wu,Tuckwell}.

\begin{table}[bht!]
\centering
\begin{tabular}{|c|c|c|c|c|c|}
\hline
 &&&&\\
 $Types$ & $ Model $    &  \b{$z_e^{(m)}$} & \b{$\lambda^2$} &  \b{$\kappa_{\lambda}^2 = \frac{1+i\omega\tau_m}{\lambda^2}$}
 \\&&&&\\
 \hline\hline
 &&&&\\
 SC &
 Standard cable
 &  \b{$z_e^{(m)} = -\frac{r_mr_e}{(r_i+r_e)(1+i\omega\tau_m)}$} 
 & \b{$\frac{r_m}{r_i+r_e}$}
 & \b{$ \frac{(r_i+r_e)(1+i\omega\tau_m)}{r_m} $}
\\ 
&
(closed-circuit)
&&&
\\&&&&\\
 \cline{1-5}
 &&&& \\
 FC &
 Frequency-dependent cable
 &  \b{$z_e^{(m)} = -\frac{r_mz_e}{(z_i+z_e)(1+i\omega\tau_m)}$} 
& \b{$\frac{r_m}{z_i+z_e}$}
 & \b{$ \frac{(z_i+z_e)(1+i\omega\tau_m)}{r_m} $}
\\ 
&
(closed-circuit)
&&&
\\&&&&\\
 \cline{1-5}
  &&&& \\
 FO
  &
 Frequency-dependent cable
  & \b{$z_e^{(m)}$}   
  & \b{$\frac{r_m}{z_i}[1+\frac{z_e^{(m)}}{r_m}(1+i\omega\tau_m)]$}
  & \b{$\frac{z_i(1+i\omega\tau_m)}{r_m[1+\frac{z_e^{(m)}}{r_m}          (1+i\omega\tau_m)]}$}
\\ 
&
 (open-circuit)
&&&
\\&&&&\\
 \cline{1-5} 
 &&&&\\
 NIC &
  Non-ideal cable     
 & \b{$ z_e^{(m)} = -\frac{\omega^2r_m\tau_m\tau_M}{[1+i\omega(\tau_m+\tau_M)][1+i\omega\tau_m]}$} 
& \b{$\frac{r_m}{z_i}[\frac{(1+i\omega\tau_m)(1+i\omega\tau_M)}{1+i\omega(\tau_m+\tau_M)}]$}
 & \b{$\frac{z_i}{r_m}~[1+ i\frac{\omega\tau_m }{1+i\omega\tau_M }] $}
\\ 
&
(closed-circuit)
&&&
\\&&&&\\
\hline
\end{tabular} 

\caption{\small Summary of dendritic cable types and parameters. 
The table gives the parameters \b{$z_e^{(m)}$}, \b{$\lambda^2$} and
\b{$\kappa_{\lambda}$} for different model types.  The standard
model (SC) is the cable model as given by Rall, Koch and
Tuckwell~\cite{Rall1962,Koch,Tuckwell}. The ``frequency-dependent
model'' (FC) correspond to a standard cable (closed circuit), but
where the parameters \b{$z_i$} and \b{$z_e$} are allowed to be
frequency dependent.  In the ``frequency-dependent open-circuit
model'' (FO), the current in the extracellular medium is
``perpendicular'' to the membrane (see Fig.~\ref{fig6}).  The
``non-ideal cable'' model (NIC) is similar to the standard model,
but the capacitance of the membrane is non-ideal, as developed
previously \cite{BedDes2008}. \b{ $z_i$} (see Eq.~\ref{eq14}) and
\b{$z_e$} are respectively the impedance per unit length of the
cytoplasm and of the extracellular medium, respectively, for FC
type models.  We write \b{$r_i$} and \b{$r_e$} when the parameters
\b{$z_s$} do not depend on frequency (SC type model).  The
parameter \b{$z_e^{(m)}$} (see Eq.~\ref{eq22}) is used in FO type
models.}

\label{table1}
\end{table}


\section{Results}

We start by generalizing the cable equations for membranes embedded
within extracellular media of arbitrarily complex electrical
properties.  Next, we consider a few specific cases and numerical
simulations.


\subsection{Generalized cable equations}

In this section, we redefine the cable equations taking into account
the presence of complex and/or heterogeneous properties of
extracellular and intracellular media. Because electrically complex or
heterogeneous media can display charge accumulation, one cannot apply
the usual (free-charge) current conservation law.  One needs to use a
more general conservation law based on the generalized current.  In
Section~\ref{gen-sec} below, we derive this generalized current
conservation law, while in Section~\ref{apl-sec}, we use this
generalized conservation law to derive the generalized cable
equations.

\subsubsection{Generalized current conservation law in
heterogeneous media} \label{gen-sec}

In this section, central to our theory, we show that the free-charge
current conservation law (\b{$\vec{j}^{~f}$}) does not apply to
systems with complex electrical properties.  Another, more general,
conservation law must be used, the generalized current conservation
law.  We derive here the conservation law for the membrane current in
arbitrarily complex media, starting from first principles.

Maxwell theory of electromagnetism postulates that the following 
relation is always valid for any medium:
\b{\textbf{
\begin{equation}
 \nabla\times\vec{H} = \vec{j}^{~f} + \frac{\partial \vec{D}}{\partial t} ~ ,
\label{eq1}
\end{equation}}}
where \b{$\vec{H}$} is the magnetic field, and \b{$\vec{j}^{~f}$}
is the current density of free charges, and \b{$\frac{\partial
\vec{D}}{\partial t}$} is the displacement current density.

We define the {\it generalized current density}
\b{$\vec{j}^{~g}$} as:
\b{
\begin{equation}
 \vec{j}^{~g}
 = \vec{j}^{~f} + \frac{\partial \vec{D}}{\partial t}
 = \vec{j}^{~f} +  \vec{j}^{~d} ~ ,
\label{eq3}
\end{equation}}
where \b{$\vec{j}^{~d}$} is the displacement current density.

It is important to note that the term
\b{$\frac{\partial\vec{D}}{\partial t} =
\varepsilon_o\frac{\partial\vec{E}}{\partial t}$} is different from
zero, even in the vacuum (assuming that the electric field varies
in time).  

\corr{The interest of using the generalized current, is that it is
  always conserved in any given volume, for any type of medium, as we
  explain below (see also Appendix~\ref{appgencurr}).}

In the case of an electric field in a homogeneous and locally
neutral medium, we have \b{$\nabla\cdot\vec{j}^{~f} =
-\frac{\partial\rho}{\partial t}^{f}=0$} because there cannot be
charge accumulation anywhere.  Because the relation
\b{$\nabla\cdot\vec{j}^{~g}=0$} applies to any type of medium, we
also have \b{$\nabla \cdot (\frac{\partial\vec{D}}{\partial
t})=0$}.  Thus, in a homogeneous locally-neutral medium, we have
two independent current conservation laws: one law applies to the
free-charge current \b{$i^{~f}$ } and another one applies to the
displacement current \b{$i^{~d}$ }.  Note that in a homogeneous
medium \b{$i^{~d}$} is not necessarily negligible, but the
application of the current conservation law on \b{$i^{~f}$} can be
done independently of the existence of \b{$i^{~d}$} because the two
laws are independent.

This is the framework assumed in the standard cable theory, in which
the extracellular medium is resistive and homogeneous, the
displacement current \b{$i^{~d}$} is negligible, and there cannot be
charge accumulation inside the dendrites nor in the extracellular
medium.  We will see below that these assumptions do not hold for
complex extracellular media. If the medium is heterogeneous, then
charge accumulation will necessarily appear in the presence of an
applied electric field.  Capacitive effects is an example of such
charge accumulation.  In such a case, the two current conservation
laws on \b{$i^{~f}$} and \b{$i^{~d}$} do not apply to every region of
space (see Appendix~\ref{appA}).  However, the generalized current
conservation on \b{$i^{~g}$} is still valid in all cases.

Thus, to derive cable equations in heterogeneous media, one must
use the generalized current conservation law, as done in the next
section.

\subsubsection{Application of the generalized conservation law to
cable equations}  \label{apl-sec}

\begin{figure}[h!] 
\centering
\includegraphics[width=8cm]{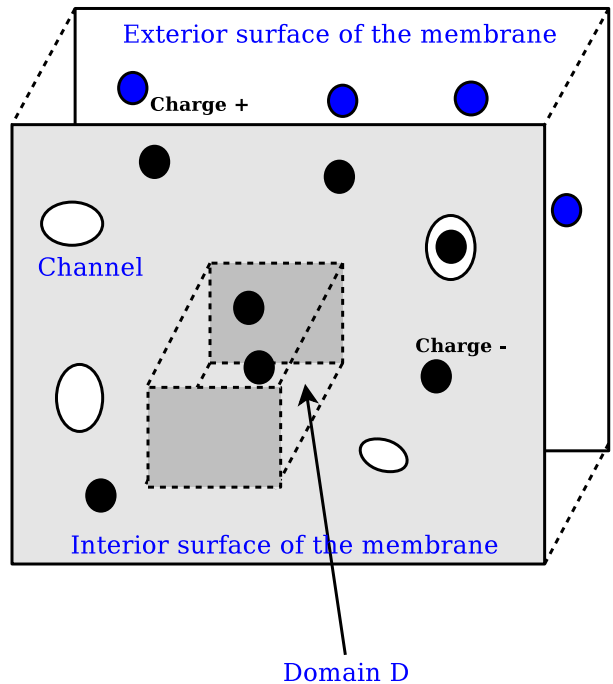}

\caption{\small \corr{(Color online)} Definition of a domain
  \b{$\mathcal{D}$} inside the cytoplasm and adjacent to the membrane.
  Due to conductance variations in the membrane, or due to charged
  currents, the total charge in domain \b{$\mathcal{D}$} varies.  One
  cannot consider that the displacement current across the surface of
  domain \b{$\mathcal{D}$} is zero, because this would be in
  contradiction with Maxwell-Gauss law (see Appendix~\ref{appA}).
  Black circles represent negative charges on the interior surface of
  the membrane, as well as in the cytoplasm, while blue circles
  indicate positive charges at the exterior side of the membrane. }
 
\label{fig1}
\end{figure}

To start, we consider a small portion of membrane surface and build
a domain in the intracellular side, which is limited by the
interior surface of the membrane, while the other surfaces of the
domain are located inside the cytoplasm (see Fig.~\ref{fig1}). 

Using such a definition, in resting conditions, the intracellular
side has an excess of negative charges, which are adjacent to the
membrane.  In such a state, we can calculate the free charge
density in this domain from Maxwell-Gauss law:
\b{
\begin{equation}
 Q(t) = \oiint\limits_{\partial \mathcal{D}} \vec{D}\cdot\hat{n}~dS = cst ~ ,
 \label{eq7}
\end{equation}}
where \b{$\partial \mathcal{D}$} is the surface of the considered
domain\footnote{Note that if the negative charges are exclusively
on the membrane (neglecting Debye layers), then the surface
integral simplifies to the side of the domain in contact with the
membrane.  In this case, the electric displacement is different from
zero only in the latter portion of the surface.}

Now, suppose that a conductance variation occurs in the domain (for
example following the opening of an ion channel).  This will induce a
charged current in domain \b{$\mathcal{D}$} and therefore, there will
be a variation of the total charge included within domain
\b{$\mathcal{D}$}, which implies a non-zero displacement current
\b{$i^{~d}$} across the surface \b{$\partial \mathcal{D}$} surrounding
domain \b{$\mathcal{D}$} (without this current, the system would be in
contradiction with Maxwell-Gauss law; see Appendix~\ref{appA}).

In such conditions, we have:
\b{
\begin{equation}
 i^{~d}=\frac{dQ}{dt}=\oiint\limits_{\partial \mathcal{D}} \frac{\partial\vec{D}}{\partial t}\cdot\hat{n}~dS \neq 0
 \label{eq8}
\end{equation}} 
Can we neglect this current to study the variations of the
membrane potential along the cable?  Because it is difficult to
give a rigorous answer to this question \cite{DesBed2012,Riera}, in
particular when \b{$i^{~d}$} is non-zero, we consider the
generalized current \b{$i^{~g}$} because this current is conserved
independently of \b{$i^{~d}$} (see previous section).  This will
allow us to treat cable equations without making any hypothesis
about charge accumulation inside or outside of the cable.

Moreover, to stay as general as possible, we include a frequency
and space dependence of the electric parameters, which will allow
us to simulate the effect of media of different electric
properties, such as capacitive or
diffusive~\cite{BedDes2006b,BedDes2009,BedDes2011a}.

In this context, the linking equations must be expressed in their
most general form~\cite{BedDes2011a}: 
\b{
\begin{equation}
\left\{
\begin{array}{ccc}
    \vec{D}(\vec{x},t) &=& \int_{-\infty}^{+\infty}
\varepsilon_i(\vec{x},t-\tau)
\vec{E}(x,\tau) ~d \tau  \\\\
    \vec{j}^{~f} (\vec{x},t)& =&\int_{-\infty}^{+\infty}
[\sigma_i^e(\vec{x},t-\tau)\vec{E}(\vec{x},\tau)
\end{array}
\right .
\label{eq9}
\end{equation}}

According to this scheme, the generalized current density
\b{$\vec{j}_i^{~g}$} inside the cytoplasm obeys:
\b{ 
\begin{equation}
\vec{j}_i^{~g} (\vec{x},t) =\int_{-\infty}^{+\infty}
[\sigma_i^e(\vec{x},t-\tau)\vec{E}(\vec{x},\tau) +
\varepsilon_i(\vec{x},t-\tau)\frac{\partial 
\vec{E}}{\partial t}(\vec{x},\tau)] ~d \tau 
\label{eq10} 
\end{equation}} 
where \b{$\sigma_i^e(\vec{x},t )$} is the intracellular electric
conductivity function and \b{$\varepsilon_i(\vec{x},t )$} is the
intracellular electric permittivity function.

\begin{figure}[h!] 
\centering
\includegraphics[width=15cm]{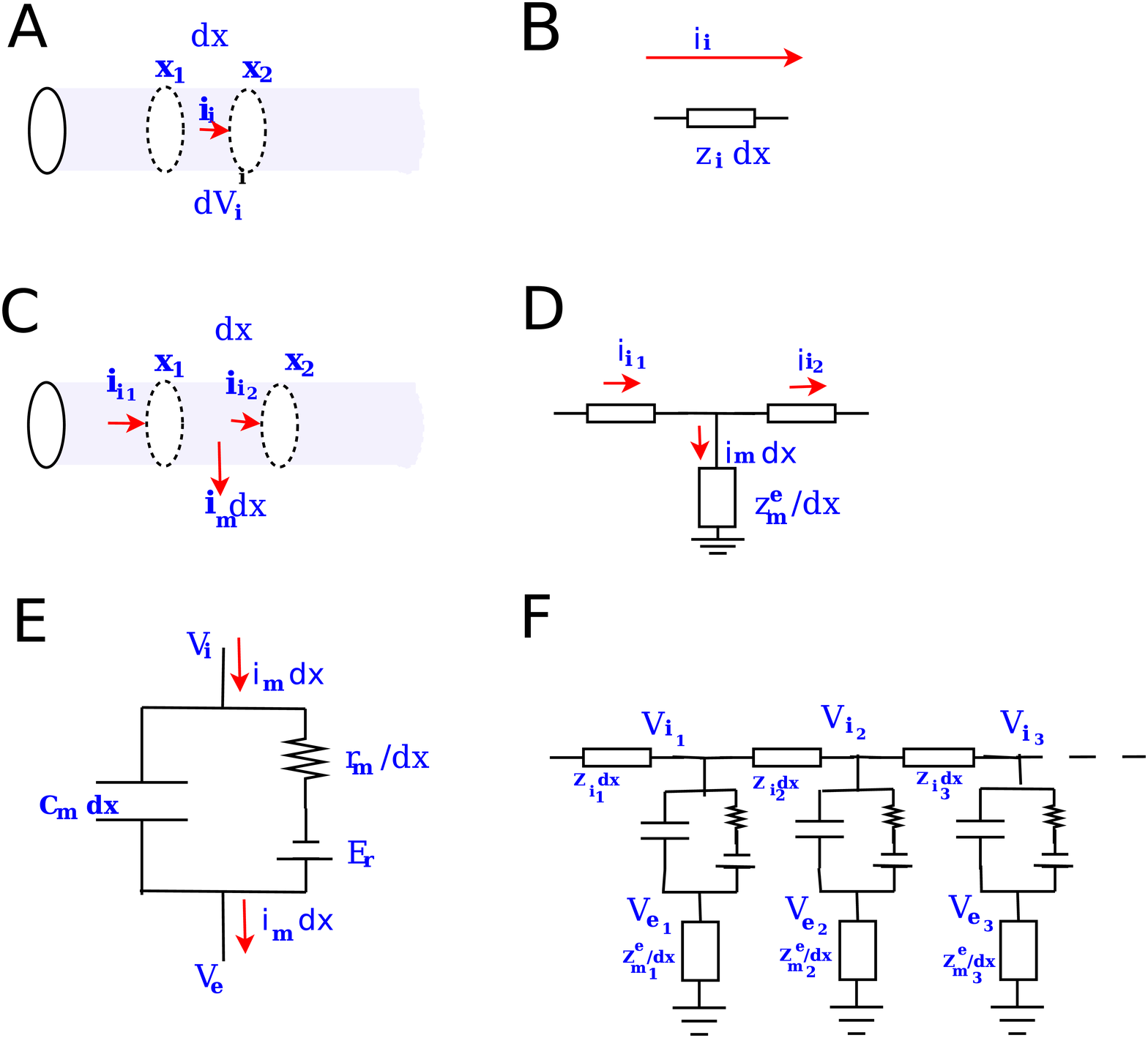}

\caption{\small \corr{(Color online)} Compartments and equivalent
  electrical circuits of the membrane and cable segments. \b{$A$} and
  \b{$C$} depict different configurations in a cable of constant
  diameter, with their respective equivalent electrical circuits shown
  in \b{$B$} and \b{$D$}. \b{$E$} is the equivalent electrical circuit
  of a membrane compartment of the cable, and \b{$F$} is the
  equivalent circuit obtained for three compartments.  \b{$V_i$} is
  the intracellular potential relative to the reference, \b{$V_e$} is
  the extracellular potential relative to the same reference,
  \b{$z_i$} is the cytoplasm impedance, \b{$r_m/dx$} and
  \b{$z_e^{(m)}/dx$} are respectively the impedances of ion channels
  and the input impedance of the extracellular medium as seen by the
  transmembrane current.  \b{$i_e$} is the output current of a cable
  element in the extracellular medium, and \b{$i_i$} is the axial
  current.  The membrane potential \b{$V_{m_j}$} equals
  \b{$V_{i_j}-V_{e_j}$} and may vary according to the position
  \b{$x_j$}. }

\label{fig2} 
\end{figure} 

The first term in the integral accounts for energy dissipation
phenomena, such as calorific dissipation (Ohm's differential law)
and diffusion phenomena.  The second term represents the effect of
charge density variations in the volume elements.  

In Fourier frequency space, \corr{Eq.~\ref{eq10}} becomes algebraic.
\b{
\begin{equation}
\vec{\j}_i^{~g}(\vec{x},\omega) =
[\sigma_i^e(\vec{x},\omega)+i\omega\varepsilon_i(\vec{x},\omega)]\vec{E}(\vec{x},\omega)
\label{eq11}
\end{equation}}
Moreover, we have \b{$\nabla\times\vec{E}=0$}, which implies
\b{$\vec{E}=-\nabla V$} because electromagnetic induction is
negligible in biological tissue (in the absence of magnetic
stimulation\footnote{In the presence of magnetic stimulation, such
as for example transcranial magnetic stimulation, we have
\cite{George1999,Landau,Purcell}: \b{$\nabla\times\vec{E} =
-\frac{\partial\vec{B}}{\partial t}\neq 0$}}).

If we now consider a one-dimensional cylindric cable of constant
radius $a$ (Fig.~\ref{fig2}A), the generalized current at a position
$x$ of the cable can be written as: \b{
\begin{equation}
 i_i^{~g}(x,\omega ) =\vec{j}_i^{~g}(x,\omega )\cdot(\pi a^2~\hat{n}) =-\pi a^2 [\sigma_i^e(x,\omega)+i\omega\varepsilon_i(x,\omega)]\cdot\frac{\partial V_i}{\partial x}(x,\omega)
\label{eq12}
\end{equation}
} where \b{$V_i$} is the intracellular voltage difference with respect
to a given reference (which can be far away).  In the following of the
text, we will call ``compartment'' a cylindric cable with constant
radius and with uniform electric parameters (see Fig.~\ref{fig0}).  It
is important to note that this compartment does not need to be
isopotential, and the membrane potential will depend on the position
on the compartment (see scheme in Fig.~\ref{fig0}). 

If we assume that the impedance (per unit length) of cytoplasm
\b{$z_i$} can be expressed as:
\b{
\begin{equation}
z_i = \frac{1}{\pi
a^2[\sigma_{i}^e(x,\omega)+i\omega\varepsilon_i(x,\omega)]} ~ ,
\label{eq13}
\end{equation}}
then the axial current can be written as:
\b{
\begin{equation}
 i_{i}^{~g}(x,\omega ) =-\frac{1}{z_i}\frac{\partial  V_{~i}}{\partial x}(x,\omega )
\label{eq14}
\end{equation}}
This expression is similar to the traditional cable
equation~\cite{Rall1995, Tuckwell, Wu}, with the exception that the
parameter \b{$z_i$} is complex (with units of
\b{$[\Omega/m]$})\footnote{This law can be simplified if one
assumes that the cytoplasm is ohmic and homogeneous (\b{$z_i=r_i$})
for negligible la permittivity.  In this case, we have:
\b{
$$
 i_i^{~g} = i_i^{~f} =\vec{j_i}^{~f}\cdot(\pi a^2~\hat{n}) =-\pi a^2\sigma_i^e \frac{\partial  V_i}{\partial x} =-\frac{1}{r_i}\frac{\partial  V_i}{\partial x}
$$
}}. In addition, the transmembrane current \b{$ i_m^{\perp}$} over a
cable length \b{$dx$} can be expressed as:
\b{
\begin{small}\begin{equation}
 i_m^{\perp}(x,t)=i_m~(x,t )~dx =
2\pi adx~[~C_m\frac{\partial V_m(x,t)}{\partial t} +\frac{\sigma_m^e}{e}(V_m(x,t)-E_m)~]
= dx~[~c_m\frac{\partial V_m(x,t)}{\partial t} +\frac{(V_m(x,t)-E_m)}{r_m}~]
\label{eq15}
\end{equation}\end{small}}
where \b{$V_m$} is the transmembrane voltage, \b{$E_m$} is the
resting membrane potential, \b{$C_m$} is the specific membrane
capacitance (in \b{$F/m^2$}), \b{$c_m$} is the membrane capacitance
per unit length (in \b{$F/m$}), \b{$\sigma_m^e$} is the electric
conductivity (in \b{$S/m$}), \b{$1/r_m$} is the linear density of
membrane conductance (in \b{$S/m$}), \b{$e$} is the membrane
thickness (in \b{$m$}) and \b{$i_m$} is the transmembrane current
per unit length (in \b{$A/m$}) (Figs~\ref{fig2}C-E). Applying the
inverse Fourier transform, we obtain:
\b{
\begin{equation}
\left \{
\begin{array}{ccccc}
i_m^{\perp}(x,0 )&=&i_m~(x,0 )~dx=~\frac{1}{r_m}[ V_m(x,0) -2\pi E_m\delta(0) ]~dx   && \omega= 0\\\\
 i_m^{\perp}(x,\omega )&=&i_m~(x,\omega )~dx  
= ~[i\omega c_m+\frac{1}{r_m} ]V_m(x,\omega )~dx
&& \omega\neq 0
\end{array}
\right .
\label{eq16}
\end{equation}}
Note that we assume here that the resting membrane potential $E_m$ does
not depend on time nor on position in the cable.  

Thus, we can see that the Fourier transform of Eq~\ref{eq14}
generates a Dirac delta function for null frequency.  In the
following of the text, we consider frequencies different from zero,
because the zero-frequency component of \b{$i_m^{\perp}$} is zero
for a signal of finite duration, which is always the case in
reality.  

In the model above, the expression of the transmembrane current is
identical to the generalized membrane current for frequencies
different from zero.  In this case, the generalized current is
given by: 
\b{
\begin{equation}
 i_m^{~g} = A \cdot j_m^{~g}= -2\pi a~dx(\sigma_m^e + i\omega\varepsilon_m)\nabla V
=2\pi a~dx(\sigma_m^e + i\omega\varepsilon_m)\frac{V_m}{l}
= dx~(\frac{1}{r_m} + i\omega c_m)V_m 
= dx~ i_m = i_m^{\perp}
\label{eq17}
\end{equation}}
where \b{$l$} is the membrane thickness and \b{$A$} is the membrane
surface.

Assuming that the charge variations inside the channels is
negligible, then the generalized current conservation law can
apply to point B in the equivalent scheme (see Fig.~\ref{fig2}~C), 
and we can write 
\b{
\begin{equation}
 i_i^{~g} (x+dx,\omega) = i_i^{~g} (x,\omega)-i_m^{~g}(x,\omega)
=i_i^{~g} (x,\omega)-i_m^{\perp}(x,\omega)
\label{eq18}
\end{equation}
}
It follows that: 
\b{
\begin{equation}
 di_i^{~g} (x,\omega) = \frac{\partial i_i^{~g}}{\partial x}~dx = -i_m^{\perp}(x,\omega)= -i_m~(x,\omega )~dx
\label{eq19}
\end{equation}
}
Using Eqs.~\ref{eq10} and \ref{eq15}, we obtain:
\b{
$$
\pi a^2 
\frac{\partial}{\partial x}[(\sigma_i^e(x,\omega)
+i\omega \varepsilon_i(x,\omega))\frac{\partial  V_i}{\partial x}(x,\omega)] 
=[i\omega c_m+\frac{1}{r_m} ]V_m(x,\omega )
$$}
Applying the partial derivative on the lefthand term, and dividing
by \b{$\pi a^2(\sigma_i^e +i\omega \varepsilon_i)$}, one obtains:
\b{
\begin{equation}
\frac{\partial^2 V_i}{\partial x^2}+\frac{1}{(\sigma_i^e
+i\omega \varepsilon_i)}\frac{\partial(\sigma_i^e
+i\omega \varepsilon_i)}{\partial x}\cdot \frac{\partial  V_i}{\partial x} 
=\frac{1}{\pi a^2(\sigma_i^e
+i\omega \varepsilon_i)}[i\omega c_m+\frac{1}{r_m} ]V_m= z_ii_m
\label{eq20}
\end{equation}}
Note that if the righthand term was zero, then this equation would
be identical to the equation describing the electric potential
outside of the sources~\cite{BedDes2004,BedDes2009,BedDes2011a},
because the \b{$\nabla $} operator equals
\b{$\hat{e}_x\frac{\partial}{\partial x} $} in one dimension, in
which case the right would be equal to \b{$ \nabla^2
V_i+\frac{\nabla \gamma_i}{\gamma_i}\cdot \nabla V_i $} where
\b{$\gamma_i = \sigma_i^e+i\omega\varepsilon_i$}.

We can simplify Eq.~\ref{eq18} if the cytoplasm is
quasi-homogeneous (assuming the scale considered is large compared
to inhomogeneities due to subcellular organelles), in which case we
can consider that the electric parameters of the cytoplasm are
independent of position x:
\b{$\sigma_i^e(x,\omega)=\sigma_i^e(\omega)$} and
\b{$\varepsilon_i(x,\omega)=\varepsilon_i(\omega)$}.  This leads to
the following expression: 
\b{
\begin{equation}
\frac{1}{z_i}\frac{\partial^2  V_{~i}}{\partial x^2}(x,\omega )=
\frac{1}{\pi a^2(\sigma_i^e
+i\omega \varepsilon_i)}\frac{\partial^2  V_{~i}}{\partial x^2}(x,\omega )
=
[i\omega c_m+\frac{1}{r_m} ]V_m(x,\omega )
\label{eq21}
\end{equation}} 
in Fourier space.

If we now assume that the extracellular medium can also be
considered as homogeneous (which will be valid at scales larger
than the typical size of the cellular elements), then we can model
the variations of the membrane potential caused by the
transmembrane current \b{$i_m^{\perp}$}.  We can model this effect
from the notion of impedance, without making any hypothesis on the
current field in the extracellular medium. In this case, one can
associate to each cable segment \b{$dx$} the specific impedance of
the extracellular medium, \b{$z_e^{(m)}$}, as seen by the
transmembrane current.  \b{$z_e^{(m)}$} has a similar physical
meaning as \b{$r_m$}, except that it is a complex number in
general. In Section~\ref{num-sec}, we will see that \b{$z_e^{(m)}$}
depends on the direction of the current field in the extracellular
medium.

Without any loss of generality, we can write in Fourier space:
\b{
\begin{equation}
 V_i(x,\omega) = V_m(x,\omega) + z_e^{(m)}(\omega)i_m(x,\omega) ~ .
\label{eq22}
\end{equation}}

By substituting this last expression in Eq.~\ref{eq19}, we obtain
\b{
$$
\frac{r_m}{z_i}[1 + \frac{z_e^{(m)}}{r_m}(1+i\omega \tau_m)]~\frac{\partial^2 V_m}{\partial x^2} =
[1+i\omega \tau_m]V_m
$$}
where \b{$\tau_m = r_m c_m$}.

Thus, we can write the system in a form similar to the standard
cable equation:
\b{
\begin{equation}
      \lambda^2\frac{\partial^2 V_m(x,\omega)}{\partial x^2} = \kappa^2 V_m(x,\omega)
\label{eq23}
\end{equation}}
where
\b{
\begin{equation}
\left \{
\begin{array}{ccc}
 \lambda^2 &=& \frac{r_m}{z_i}\cdot[1+\frac{z_e^{(m)}}{r_m}(1+ i\omega \tau_m)]=\frac{r_m}{\bar{z}_i} \\\\
 \kappa^2 & = & 1+i\omega\tau_m 
\end{array}
\right .
\label{eq24}
\end{equation}}
for a cylindric compartment (see Eq.~\ref{eqb8} in
Appendix~\ref{appB}).  It follows that the general solution of
this equation in Fourier space \b{$\omega \neq 0$} is given by:
\b{
\begin{equation}
 V_m(x,\omega) = A^{+}(\omega)e^{\frac{\kappa (l-x)}{\lambda}}
+A^{-}(\omega)e^{\frac{-\kappa (l-x)}{\lambda}}
\label{eq25}
\end{equation}}
for each cylindric compartment of length \b{$l$} and with constant
diameter (see Fig.~\ref{fig0} for a definition of coordinates). 
For a given frequency, we have a second order differential equation
with constant coefficients.

In general, one can apply Eq.~\ref{eq25} for different cylindric
compartments, as in Fig.~\ref{fig2}F.  In this case, one must adjust
the different compartments to their specific limit conditions
(continuity of \b{$V_m$} and of the current
\b{$i_i^{~g}=-\frac{1}{\bar{z}_i}\frac{\partial V_m}{\partial x}$}
(see Eq.~\ref{eqb4} in Appendix~\ref{appB}).

Note that Eq.~\ref{eq25} is exact for a cylindric compartment of
constant diameter.  Thus, it is possible to use this property to
simulate exactly the full cylindric compartment as a continuum with no
need of spatial discretization into segments, as usually done in
numerical simulators.  This is only possible if the cylindric
compartment has a constant diameter.  This leads to an efficient
method to simulate the cable equations.  We will refer to this
approach as ``continuous compartment'' in the following.

As mentioned above, the mathematical forms of Eqs.~\ref{eq23} and
\ref{eq25} are identical to that of the standard cable model, but with
different definitions of \b{$\lambda$}.  Thus, we directly see that
the nature of the extracellular medium will change the value of these
parameters, which become frequency dependent.  In particular, we see
from Eq.~\ref{eq23} that changing these parameters will impact on the
spatial profile of the variations of \b{$V_m$}, if the frequency
dependence of the ratio \b{$\kappa_{\lambda} =\frac{\kappa}{\lambda}$}
is affected by the nature of the medium.  Thus, experimental
measurement of the spatial variations of \b{$V_m$} will be able to
identify effects of the extracellular impedance only if the ratio
\b{$\kappa_{\lambda}$} is affected.

In the next section, we derive expressions to calculate the input
impedance \b{$Z_{in}(P)= \frac{V_m(P,\omega)}{i_i(P,\omega)}$} and the
transfer function of the transmembrane voltage \b{$F_T(\omega)=
  \frac{V_m(P_{b},\omega)}{V_m(P_{a},\omega)}$} between two positions
in the cable (as a function of the ratio \b{$\kappa_{\lambda}$}).
Later in Section~\ref{num-sec}, we will see that it is necessary to
know these quantities to calculate the spatial variation of \b{$V_m$}
and compare the standard model with the cable model embedded into
complex extracellular media.


\subsubsection{Method to solve the generalized cable}

In this section, we present the theoretical expressions which will
allow us to calculate the input impedances needed for computing the
membrane voltage on a cable with varying diameter.  We consider the
input impedance of the membrane, as well as the impedance of the
extracellular medium, both of which are needed to calculate the
spatial profile of the \b{$V_m$} in a given cable segment.

We proceed according to the following steps:

1. In the previous section, we saw that it is necessary to
calculate the ratio \b{$Z_{in}(P)=
\frac{V_m(P,\omega)}{i_i(P,\omega)}$} at the position of the
current source, to calculate the \b{$V_m$} produced at that point. 
\begin{figure}[bh!] 
\centering
\includegraphics[width=10cm]{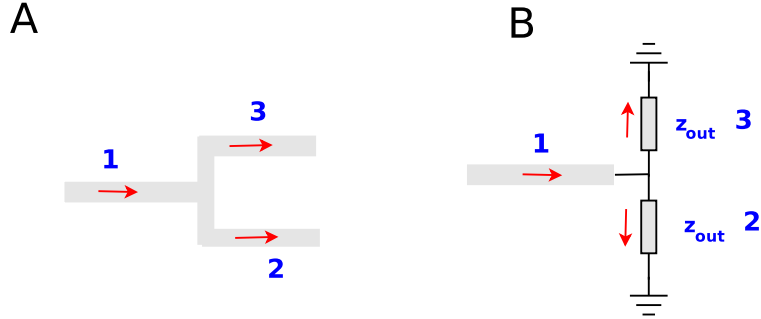}

\caption{\small \corr{(Color online)} Branching cables.  The panels A
  and B respectively represent a branched cable where a dendrite
  separates into two daughter branches, and its equivalent electrical
  circuit.  The equivalent impedance of segment 1 is equal to the
  input impedances of segments 2 and 3 (\b{$z_{out~2}$} and
  \b{$z_{out~3}$} ) taken in parallel.}

\label{fig4}
\end{figure}

One strategy is, in a first step, to separate the cable into a
series of continuous compartments of constant diameter, where
parameters \b{$a$} (Eq.~\ref{eq12}), \b{$z_i$} (Eq.~\ref{eq14}),
\b{$r_m$} (Eq.~\ref{eq15}) and \b{$z_e^{(m)}$} (Eq~\ref{eq22}) are
constant and specific to each compartment.  In a second step, one
calculates the (transmembrane) input impedance
\b{$Z_{in}^{n+1}=\frac{V_m(0)}{i_i(0)}$} at the begin of each
compartment by taking into account the auxiliary impedance at the
end of this compartment, \b{$Z_a =Z_{out}^{n+1}=
\frac{V_m(l_{n+1})}{i_i(l_{n+1})} = Z_{in}^{n}  $} (see
Fig.~\ref{fig0}) if there is no branching point.  At the branching
points, the auxiliary impedances are simply equal to the
equivalent input impedance of $n$ dendritic branches in parallel
(where $n$ is the number of ``daughter'' branches; 
see Fig.~\ref{fig4}). Thus, because the input impedance at one end
is equal to the input impedance of the other compartment connected
to this end, one obtains a recursive relation (see Eq.~\ref{eqb9}
in Appendix~\ref{appB}):
\b{
\begin{equation}
Z_{in}^{n+1}[Z_{in}^{n}] = \frac{\bar{z}_{i_n}}{\kappa_{{\lambda}_n}}~
\frac{(\kappa_{{\lambda}_n}Z_{in}^{n}+\bar{z}_{i_n} )~e^{2\kappa_{{\lambda}_n}l_n} +( \kappa_{{\lambda}_n}Z_{in}^{n}-\bar{z}_{i_n})}{(\kappa_{{\lambda}_n}Z_{in}^{n}+\bar{z}_{i_n} )~e^{2\kappa_{{\lambda}_n}l_n} -(\kappa_{{\lambda}_n}Z_{in}^{n}-\bar{z}_{i_n}) }
\label{eq26}
\end{equation}
}
where
\b{
$$
\bar{z}_i = \frac{z_i}{1+\frac{z_e^{(m)}}{r_m}(1+i\omega\tau_m)}
$$
}
Thus, we can write \b{$$Z_{in}^{n+1}=
F~[Z_{in}^{n}~;~\bar{z}_{i_{n}},\kappa_{\lambda_n},l_n  ]$$} This leads
to the following expression to relate the first to the \b{$n^{th}$}
segment:
\b{\begin{equation}
Z_{in}^{n+1}= F~[...F~[F~[Z_{in}^{1}~;~\bar{z}_{i_{1}},\kappa_{\lambda_1},l_1  ];~\bar{z}_{i_{2}},\kappa_{\lambda_2},l_2]...;~\bar{z}_{i_{n}},\kappa_{\lambda_n},l_n]
\label{eq27}
\end{equation}}

Note that this algorithm is a generalization of that used to calculate
the equivalent resistance for resistances in series.  Indeed, for
resistance in series we have \b{$r_{eq}=F(...F(r_1;r_2);r_n)$} where
\b{$F(r_{a};r_{b})=r_a+r_b$}.  The difference between this recurrence
function and that of Eq.~\ref{eq29} essentially comes from the fact
that there is no current leak in a resistance, while there is
one in a dendritic compartment.

2. To calculate the profile of \b{$V_m$} along the cable, one must use
the spatial transfer function \b{$
  \frac{V_m(P_{n+1},\omega)}{V_m(P_{n},\omega)}$} on a continuous
cylindric compartment of arbitrary length, and calculate the product
of the transfer functions between each connected compartment.  This
leads to (see Appendix~\ref{appC} and Eq.~\ref{eqc3}): \b{
\begin{equation}
F_T(l,\omega ; Z_{out}^n) = \frac{\kappa_{\lambda}Z_{out}^n}{\kappa_{\lambda}Z_{out}^n~cosh (\kappa_{\lambda}l) +\bar{z}_i sinh(\kappa_{\lambda}l)}
\label{eq28}
\end{equation}
}
\b{
\begin{equation}
 \frac{V_m(P_n,\omega)}{V_m(P_1,\omega)}
=\prod_{i=1}^{n-1}
\frac{V_m(P_{i+1},\omega)}{V_m(P_i,\omega)}
\label{eq29}
\end{equation}
}

3. To evaluate \b{$z_{proximal}$} we must calculate the first
impedance \b{$Z_{in}^1$} which enters the recursive relation
\ref{eq28}. This impedance corresponds to the impedance
of the soma, which is given by: 
\b{
\begin{equation}
Z_{in}^1=Z_s +Z_{cs} ,
\label{eq30}
\end{equation}}
where \b{$Z_s$} is the soma membrane impedance and 
\b{$Z_{cs}$} is the cytoplasm impedance inside the soma. This relation is 
obtained under the hypothesis that the soma is isopotential, and the 
application of the generalized current conservation law implies
\b{$i^{~g}=\frac{V_i-V_e}{Z_s +Z_{cs}}\approx\frac{V_m}{Z_s +Z_{cs}} $} 
where \b{$V_i$} and \b{$V_e$} are the electric potentials at both sides of
the membrane, inside and outside, respectively relative to a reference 
located far-away. 

The impedance of the bilipidic membrane is approximated by a parallel
\b{RC} circuit where \b{$R=R_m$} is the resistance and
\b{$\tau_m=R_mC_m$} is the membrane time constant.  Thus,
\b{$Z_{in}^1$} can be written as: \b{
\begin{equation}
Z_{in}^1 =  Z_s + Z_{cs} =\frac{R_m }{1+i\omega \tau_m}+Z_{cs}
\label{eq31}
\end{equation}
} 

Finally, to evaluate \b{$z_{distal}$}, we use the ``sealed end''
boundary condition \b{$Z_{in}^1=\infty$}.   In this condition, we
have \b{$Z_{in}^2 =
\frac{\bar{z}_{i_1}}{\kappa_{\lambda_1}}~coth(\kappa_{\lambda_1}l_1)$}
(see Eq.~\ref{eq26}).  In the case of a single dendritic branch,
we can write:
\b{
\begin{equation}
Z_{in}^{distal} = 
\frac{\bar{z}_{i}}{\kappa_{\lambda}}~coth(\kappa_{\lambda}l) ~ ,
\end{equation} \label{eq32}} 
where \b{$l$} is the total length of the cable.

In the next section, we turn to numerical simulations to
investigate passive cable properties in the presence of
complex media.  We consider the most general case, where both the
impedance of the extracellular medium and that of cytoplasm can be
frequency dependent, and determine the respective impact on the
spatial profile and frequency content of the transmembrane voltage
at the level of the proximal and distal ends of the cable.

\subsection{Numerical simulations}
\label{num-sec}

The goal of the numerical simulations is here to show how the
physical nature of extracellular and intracellular media can
influence the spatial and frequency profiles of the transmembrane
potential.  We present simulations of a ``continuous ball and
stick'' model, which consists of a continuous cylindric compartment
(described by Eq.~\ref{eq25}), connected to a spherical soma.  In
this case, the impedance \b{$Z_a$} of the continuous cylindric
compartment is the soma impedance (see Fig.~\ref{fig0}).  We do not
investigate here the effect of complex dendritic structures, which
is left for future studies.  Note that what we call a ``continuous
cylindric compartment'' actually represents an infinite number of
compartments each represented by a resistance in series with a
parallel RC circuit (see Fig.~\ref{fig2}F).

In a first step, we list the different types of models of
intracellular and extracellular media that were used.  In a second
step, we present the results of numerical simulations.

\subsubsection{Different types of cable models}

We now explain the parameters used for the simulations of
the cable presented in Section~\ref{AnNum}. 

Because the cable equation (Eq.~\ref{eq23}) is completely determined
by the value of \b{$\kappa_{\lambda}$} for a given frequency, the
spatial and frequency profiles of the transmembrane voltage are
completely determined if the geometry and boundary conditions are set.
And because \b{$\kappa_{\lambda}$} is a function of 4 parameters
(\b{$r_m,\tau_m,z_i, z_e^{(m)}$})(Eq.~\ref{eq24}) for a given
frequency, we have a four-dimensional parameter space where the two
last parameters (\b{$z_i, z_e^{(m)}$}) can be frequency dependent.  We
will limit our exploration of this parameter space by only varying the
physical nature of these impedances for realistic values of \b{$r_m$}
and \b{$\tau_m$}, because the influence of these parameters has been
largely characterized in previous studies \cite{Rall1962,Koch,Wu}.
Furthermore, with \b{$\tau_m$} and \b{$\omega $} fixed, the relation
\b{$\kappa_{\lambda}=\frac{1+i\omega\tau_m}{\lambda}$} depends only on
\b{$\lambda$}, and thus, like the classic studies on cable equations,
we will use this parameter as a main determinant of the cable
properties.

We will explore the generalized cable equations by considering
several typical cases:

\begin{figure}[h!] 
\centering
\includegraphics[width=10cm]{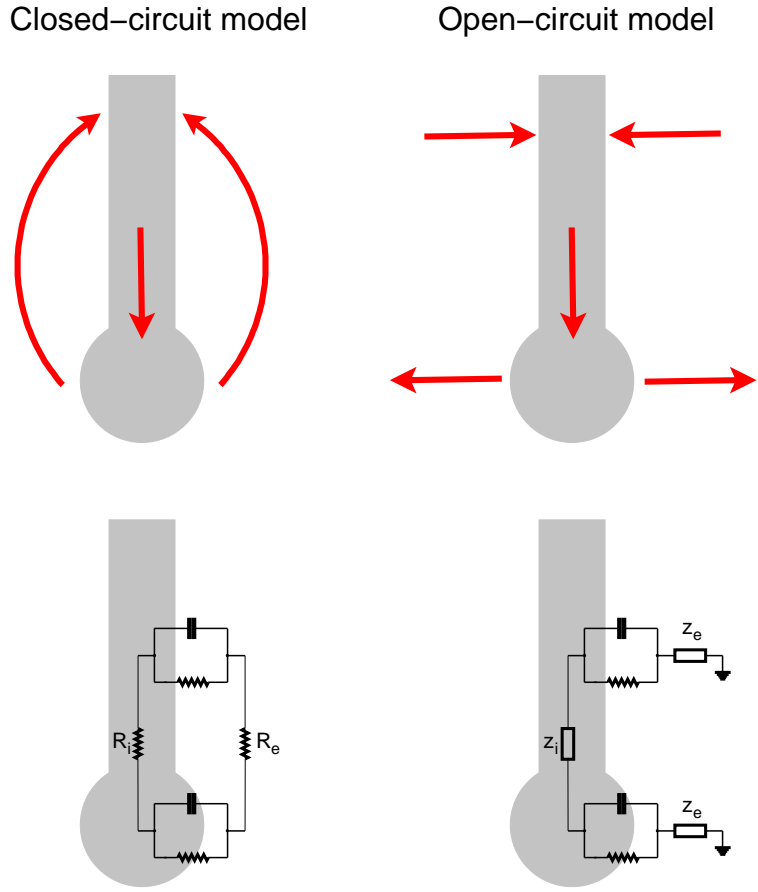}

\caption{\small \corr{(Color online)} Two different cable models for
  neurons.  Left: {\it Closed-circuit model}.  This is the standard
  cable model which forms a closed system (all inward and outward
  currents are balanced) and can be described by an equivalent circuit
  (bottom; shown here for a two-compartment model; $R_e$ and $R_i$ are
  the extracellular and intracellular resistances, respectively).  In
  this model, the current flows parallel to the neuron.  Right: {\it
    Open-circuit model}.  In this more general model, the current is
  allowed to flow between neighboring neurons, or between the neuron
  and extracellular space, with no necessary condition of local
  balance (top).  In this case, the neuron is modeled by an open
  circuit (bottom), and the current flows ``perpendicular'' to the
  membrane. The equivalent circuit is modeled more generally with
  impedances ($Z_e$ extracellular, $Z_i$ intracellular)}

\label{fig5} 
\end{figure}

\paragraph*{Standard cable model}

\begin{figure}[bht!] 
\centering
\includegraphics[width=13cm]{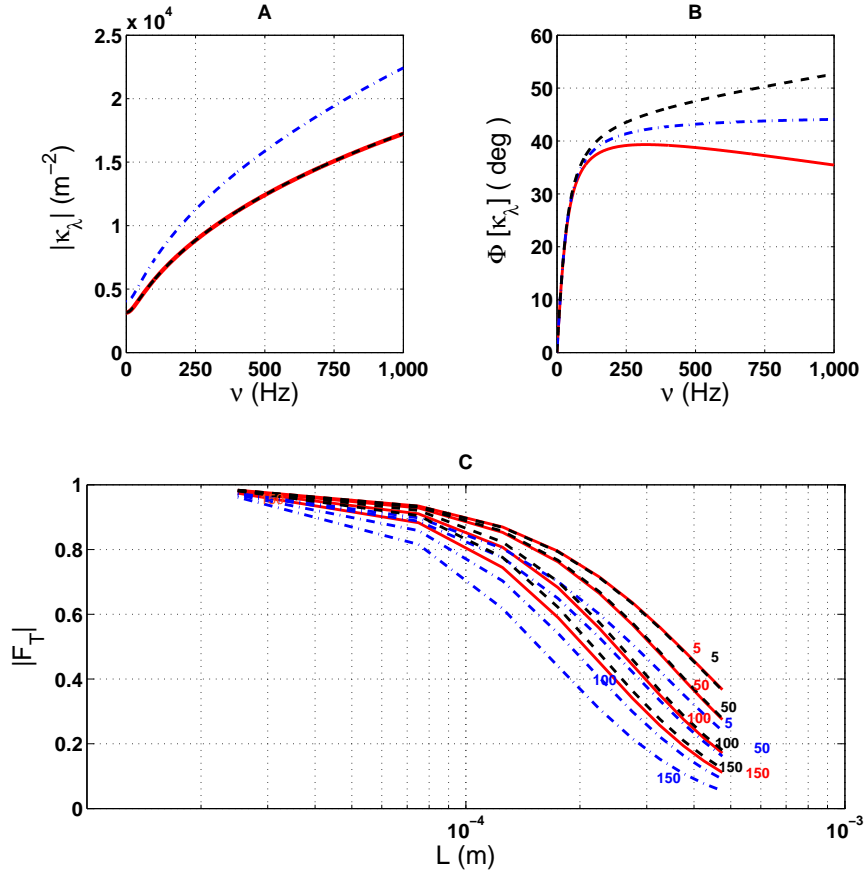}

\caption{\small \corr{(Color online)} Spatial and frequency profile of
  the membrane potential in the cable model with resistive media.  A
  and B respectively show the modulus \b{$|\kappa_{\lambda}|$} and the
  phase \b{$\Phi[\kappa_{\lambda}]$} of \b{$\kappa_{\lambda}$} as a
  function of frequency \b{$\nu$} for a continuous ball-and-stick
  model.  C.  Modulus of the transfer function \b{$|F_T|$} as a
  function of distance \b{$L$} in the dendritic compartment, for
  frequencies equal to 5, 50, 100 and 150 Hz (see corresponding
  frequencies in A and B).  The \corr{blue curves in $-\cdot -$}
  correspond to a standard cable model (FC, closed-circuit), with
  \b{$r_i = 28\times 10^9 ~\Omega/m$} and \b{$r_e = 18\times 10^9
    ~\Omega/m$} .  The \corr{red} curves correspond to the same model
  but in an open-circuit configuration (FO model), with \b{$r_i =
    28\times 10^9 ~\Omega/m$} and \b{$z_e^{(m)}=0.01~\tau_m/2\pi
    aC_m=0.4\times10^3~\Omega.m$}.  The \corr{black curves in $--$}
  show a non-ideal cable (NIC) model with \b{$\tau_M=0.01\tau_m $},
  \b{$r_i = 28\times 10^9 ~\Omega/m$} and \b{$r_e = 0 ~\Omega/m$}.}

\label{fig6}
\end{figure}

The first type of model that we will consider is the ``standard
cable model'' (model SC in Table~\ref{table1}), identical to that
considered by Rall, Koch and Tuckwell \cite{Rall1962,Koch,Tuckwell}. 
In this model, the neuron is a closed system, where the inward and
outward currents are balanced, forming a closed circuit (see
Fig.~\ref{fig6}, left).  The extracellular current flows {\it
parallel} to the dendrite, as noted previously \cite{Tuckwell}. 
This model is equivalent to consider that the field produced by the
neuron corresponds to an electric dipole configuration.  In
addition, this model considers that the extracellular medium is
resistive, or in other words, that the extracellular impedance is a
constant.

In this standard model, the extracellular impedance \b{$z_e^{(m)}$}
is either zero (no extracellular resistivity) as in Rall's and
Koch's formulations~\cite{Rall1962, Rall1995,Koch}, or is equal to
a constant, which is equivalent to model the extracellular medium
by a resistance, as in other
formulations~\cite{Tuckwell,Rieracable}. Besides its physical non-sense
(the extracellular medium considered as a supraconductor), using a
zero-resistance is usually justified from the fact that the
extracellular resistivity is much smaller than the membrane
impedance.  We will see that this justification does not hold if
the medium is frequency dependent, in which case for some frequency
range the extracellular resistivity may be determinant.  Thus, to
obtain the general expression of \b{$\lambda$} and
\b{$\kappa_{\lambda}$} for the standard model, we set \b{$z_e^{(m)}
= -\frac{r_mr_e}{(r_i+r_e)(1+i\omega\tau_m)}$} in Eq.~\ref{eq24}
(see Table~\ref{table1}).

\paragraph*{Frequency-dependent cable model}
\label{freqdepcab}

The second type of model is an extension of the standard model, where
the intracellular and extracellular impedances (\b{$z_i$} and
\b{$z_e^{(m)}$}, respectively) are allowed to depend on frequency.
This ``frequency-dependent cable model'' (model FC in
Table~\ref{table1}) can account for example for a neuron embedded in
capacitive or diffusive\footnote{A medium is said to be ``diffusive''
  when ionic diffusion is non-negligible in the presence of an
  electric field.} extracellular media, or if the intracellular medium
has such properties, or both.  In such cases, the appropriate
frequency-dependent profiles for the impedances must be used.

In this frequency-dependent model, if \b{$\tau_m$} is fixed, the
quantity \b{$z_e+z_i$} completely determines the spatial and
frequency profiles of the V$_m$, and how they deviate from the
standard model (see Table~\ref{table1}).  To explore the effect of
the impedances \b{$z_i+z_e$}, we consider three typical cases:
``resistive'', ``capacitive'' (which is in fact resistive and
capacitive in parallel) and ``diffusive'' (which is equivalent to a
Warburg type impedance). Such impedances have also been considered
in previous studies \cite{BedDes2010,BedDes2009,Poz}.  

Note that, in order to simulate the standard model, one must
necessarily assume that the real part of \b{$z_e^{(m)}$} is
negative\footnote{For example, if \b{$z_e =r_e$} and \b{$z_i=r_i$},
then we have \b{$R(z_e^{(m)}) =
-\frac{r_mr_e}{r_i+r_e}\frac{1}{1+\omega^2\tau_m^2}< 0$}.}, which
implies that \b{$z_e^{(m)}$} is not a passive impedance per unit
length, but is active, and thus requires a source of energy, as
pointed previously \cite{Conciauro, Mauro}. This point will be
further considered in the Discussion.  

\begin{figure}[bht!] 
\centering
\includegraphics[width=13cm]{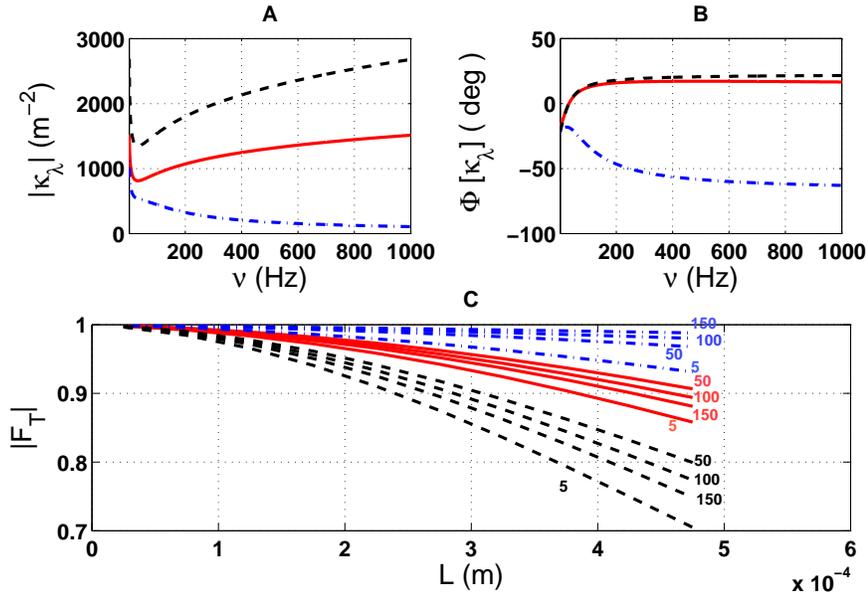}

\caption{\small \corr{(Color online)} Spatial and frequency profiles
  of the membrane potential for a model with resistive extracellular
  medium and diffusive cytoplasm.  A and B: modulus
  \b{$|\kappa_{\lambda}|$} and phase \b{$\Phi[\kappa_{\lambda}]$} of
  \b{$\kappa_{\lambda}$} as a function of frequency, for a continuous
  ball-and-stick model.  C.  Modulus of the transfer function
  \b{$|F_T|$} as a function of distance for different frequencies
  (same arrangement as Fig.~\ref{fig6}).  The red curves correspond to
  a model with zero extracellular resistance.  The blue curves
  \corr{($-\cdot -$)} show models with open-circuit configuration (FO
  model with \b{$z_e^{(m)}=0.5~\tau_m/2\pi aC_m =
    20\times10^3~\Omega.m$}) and diffusive cytoplasm (\b{$z_i =
    \frac{28\times 10^9}{(1+i)\sqrt{w}} ~\Omega/m$}).  The black
  curves \corr{($--$)} show the same model with closed-circuit
  configuration with a resistive extracellular medium (FC model
  \b{$r_e = 18\times 10^9 ~\Omega/m$}).  Note that for the FC model,
  \b{$|F_T|$} progressively increases from 5 to 50~Hz, then decreases
  between 50 and 100~Hz.}

\label{fig7}
\end{figure}

\paragraph*{Open-circuit model}

In a third type of model, the ``Open-circuit'' model (FO in
Table~\ref{table1}), we use a different approach.  Instead of
considering the neuron as a closed system, where all outward currents
must return to the neuron, we make no hypothesis about the return
currents, and allow for example that neighboring neurons exchange
currents\footnote{This will be the case for example if two neighboring
  dendrites have current sources of opposite sign, there will be a
  direct current flow between them.  If they belong to different
  neurons, this configuration necessarily requires an open-circuit
  model to be accounted for.}.  In this case, one does not need to
describe each neuron by a closed circuit, but all neurons are open
circuits are are connected together (through the extracellular space).
Figure~\ref{fig6} shows the current fluxes of the two models, the
standard model is a closed circuit where the outward currents loop
into the inward currents (Fig.~\ref{fig6}A), while in the open-circuit
model, all currents are exchanged with the surrounding medium
(Fig.~\ref{fig6}B).  These two models correspond to different
equivalent circuits \corr{(Fig.~\ref{figF1} in Appendix~\ref{appF}).}

\begin{figure}[bht!] 
\centering
\includegraphics[width=13cm]{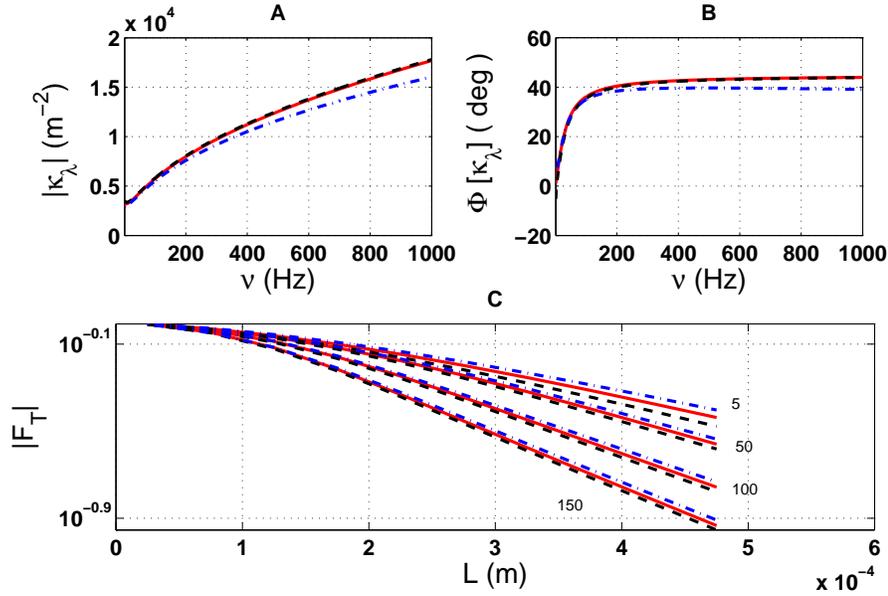}

\caption{\small \corr{(Color online)} Spatial and frequency profiles
  for a model with resistive cytoplasm and diffusive extracellular
  medium.  Same arrangement of panels as for Figs.~\ref{fig6} and
  \ref{fig7}, but for different media.  The black curves \corr{($--$)}
  show the behavior of a closed-circuit (FC) type model with resistive
  cytoplasm (\b{$r_i = 28\times 10^9 ~\Omega/m$}) and diffusive
  extracellular space with Warburg impedance (\b{$z_e = \frac{18\times
      10^9}{(1+i)\sqrt{w}} ~\Omega/m $}). The red curves correspond to
  a closed-circuit (FC) type model with \b{$ z_i = 28\times 10^9
    ~\Omega/m $} and \b{$z_e = 0 ~\Omega/m $}.  The blue curves
  \corr{($-\cdot -$)} correspond to an open-circuit (FO) type model
  (\b{$r_i = 28\times 10^9 ~\Omega/m$},
  \b{$z_e^{(m)}=\frac{\tau_m}{2\pi aC_m}\frac{0.5}{(1+i)\sqrt{w}} =
    \frac{20\times 10^3}{(1+i)\sqrt{w}}~\Omega.m$}).}

\label{fig8}
\end{figure}

Note that the Open-circuit cable model is practically equivalent to
the traditional (closed-circuit) cable model for an isolated
neuron, if the impedance of the extracellular medium is negligible
compared to the membrane impedance.  Indeed, if \b{$z_e^{(m)}$}
and \b{$z_e$} tend to \b{$0$}, then we have
(see Table~\ref{table1}):
\b{
\begin{equation}
 \lim_{\b{z_e^{(m)}}\rightarrow 0}\lambda_{FO}^2
= \frac{r_m}{z_i} =\lim_{z_e \rightarrow 0}\lambda_{FC}^2
\label{eq35}
\end{equation}}

Similar to the frequency-dependent cable model, we will consider
the three types of impedances discussed above (resistive,
capacitive and diffusive) in the simulations of the Open-circuit
model.  In this case, we separately consider the two quantities
\b{$z_i$} and \b{$z_e^{(m)}$} because these two parameters directly
determine the value of \b{$\lambda$} in models of FO type (see
Table~\ref{table1}).  Note that in the Open-circuit model, the real
part of \b{$z_e^{(m)}$} is always positive, so there is no need of
any additional energy source (see Discussion).

\begin{figure}[bht!] 
\centering
\includegraphics[width=13cm]{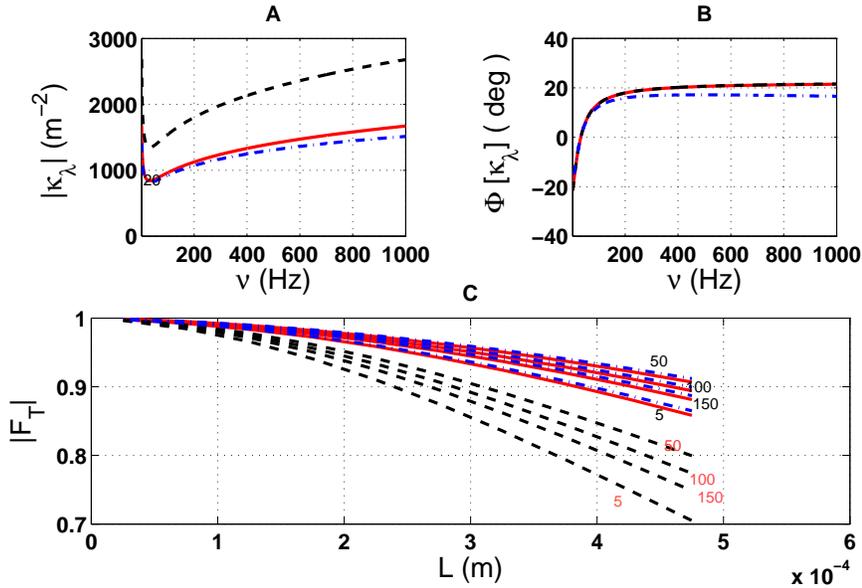}

\caption{\small \corr{(Color online)} Spatial and frequency profiles
  for fully diffusive cable models.  Same arrangement of panels as for
  Figs.~\ref{fig6}--\ref{fig8}, but using a continuous ball-and-stick
  model where both cytoplasmic and extracellular impedances are of
  diffusive (Warburg) type.  The black curves \corr{($--$)} correspond
  to a closed-circuit (FC) type model with \b{$ z_i = \frac{28\times
      10^9}{(1+i)\sqrt{w}} ~\Omega/m $} and \b{$z_e = \frac{18\times
      10^9}{(1+i)\sqrt{w}} ~\Omega/m $}. The red curves correspond to
  a closed-circuit (FC) type model with \b{$ z_i = \frac{28\times
      10^9}{(1+i)\sqrt{w}} ~\Omega/m $} and \b{$z_e = 0 ~\Omega/m $}.
  The blue curves \corr{($-\cdot -$)} correspond to a closed-circuit
  (FO) type model with \b{$ z_i = \frac{28\times 10^9}{(1+i)\sqrt{w}}
    ~\Omega/m $} and \b{$z_e^{(m)}= \frac{20\times
      10^3}{(1+i)\sqrt{w}}~\Omega.m$}.  Note that for both types of
  models (FO and FC), \b{$|F_T|$} increases between 5 and 50~Hz, then
  decreases between 50 and 100~Hz.}

\label{fig9}
\end{figure}

\paragraph*{Non-ideal cable model}

The fourth type of model considered here is the ``non-ideal cable
model'' introduced previously~\cite{BedDes2008}.  This model
postulated that the membrane capacitance is non-ideal, \corr{through}
the use of an additional resistance at the arms of the capacitor; this
resistance models the fact that there is some inertia time to charge
movement (or equivalently, a friction).  Such a non-ideal capacitance
resulted in a shallower frequency scaling, that is a higher capacity
of the dendritic tree to propagate high-frequency
events~\cite{BedDes2008}.  Note that in this model, the extracellular
medium is modeled as a resistance, so in this respect, the non-ideal
cable model is equivalent to the standard model.  Mathematically, the
non-ideal cable appears through the use of \b{$z_e^{(m)}$} (see
Table~\ref{table1}), which can therefore be viewed as a particular
case of an influence of the extracellular medium on cable properties.
Indeed, the non-ideal cable can be shown to be equivalent to -- or a
particular case of -- the open-circuit model, where the \b{$V_m$}
corresponds to \b{$V_i$} with a far-away reference (see
Appendix~\ref{appF}).  We keep this model here for comparison.

\subsection{Simulation of the different models}
\label{AnNum}

In this section, we present the results of numerical simulations of
the models presented in the previous section (see Methods). The goal
of these simulations is not to be exhaustive in considering all
possible combinations of models, but present a few typical
configurations.  The central question is whether the nature of the
extracellular medium can have determinant impact on cable properties,
and for what type of configuration or parameter values does it
happen~?

\begin{figure}[bht!] 
\centering
\includegraphics[width=10cm]{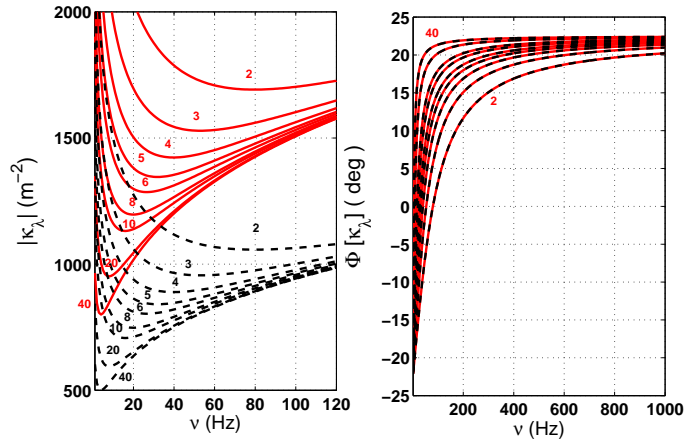}

\caption{ \corr{(Color online)} Parameter \b{$\kappa_{\lambda}$} as a
  function of frequency for fully diffusive models. The black curves
  \corr{$--$} correspond to FO and the red curves to FC type models
  with a time constant of \b{$\tau_m=2,3,4,5,6,8,10,20$} and \b{$40
    ~ms$} . The FC type model was with \b{$ z_i = \frac{28\times
      10^9}{(1+i)\sqrt{w}} ~\Omega/m $} and \b{$z_e = \frac{18\times
      10^9}{(1+i)\sqrt{w}} ~\Omega/m$} .  For the FO type model,
  \b{$z_i = \frac{28\times 10^9}{(1+i)\sqrt{w}} ~\Omega/m $} and
  \b{$z_e^{(m)}=\frac{\tau_m}{2\pi a C_m}\frac{0.5}{(1+i)\sqrt{w}} =
    \frac{20\times 10^3}{(1+i)\sqrt{w}}~\Omega.m$} (see
  Table~\ref{table2} for the corresponding resonance frequencies).}

\label{fig10}
\end{figure}

\subsubsection*{Analysis of the spatial profiles of V$_m$
variations} \label{anal-sec}

In this section, we investigate analytically and numerically different
particular cases of extracellular and intracellular media to determine
how the nature of these media affects the spatial and frequency
profile of the membrane potential.  We consider the transfer functions
as defined in Table~\ref{table1}.  The analyses presented here are
limited to a ball-and-stick model, which allows a better
interpretation of the effect of the physical nature of the different
media.  The effect of complex dendritic tree morphology will be the
subject of a future study.  To compare the results from the different
models, we have considered models with identical geometry (see Methods
for parameters).

\paragraph*{Resistive models}

We first considered the ``standard model'' with resistive
intracellular and extracellular media, as well as the non-ideal cable
model \cite{BedDes2008}.  In Figure~\ref{fig7}, we can see that the
nature of the cable model (closed-circuit or open-circuit; non-ideal)
influences the modulus and the phase of \b{$\kappa_{\lambda}$}, as
well as the spatial profile of the transfer function \b{$|F_T|$}. The
modulus of the transfer function depends more strongly on frequency in
the FC model compared to the two other cases (Fig.\ref{fig7}C), as
observed previously \cite{BedDes2008}. Note that the parameters of the
FO and NIC models were chosen such that they are equivalent (see
Appendix~\ref{appA}).

\paragraph*{Capacitive models}

Next, we considered models where the cytoplasm and extracellular
medium are both of capacitive (RC-circuit) type.  Note that we
considered capacitive effects without ionic diffusion, because if both
are combined, the resulting impedance is of Warburg type.  This type
of model will be considered next.  With purely capacitive media, we
observed effects very similar to the resistive model shown in
Fig.~\ref{fig7}, with slight differences only visible for large
frequencies (greater than about \b{$200~Hz$} (not shown).  The small
dimension of organelles (\b{$<<1~\mu m^2$}) within cells, as well as
the distance between neighboring cells (\b{$\sim 30~nm$} on average)
\cite{Braitenberg,Nicho2005} imply that the capacitance values of the
media are necessary small compared to the membrane capacitance, and
thus the purely capacitive effects (without diffusion) are likely to
be negligible.

\begin{table}[bht!]
\centering
\begin{tabular}{|c|c|c|c|}
\hline
 \b{$\tau_m~(ms)$}  & \b{$\nu_r~(Hz)$}  \\
  \cline{1-2} 
 2  &  83 \\
 3  &  54 \\
 4  &  40 \\
 5  &  30 \\
 6  &  25 \\
 8  &  20 \\
 10 &  18 \\
 20 &   8 \\
\hline
\end{tabular} 

\caption{Resonance frequencies of fully diffusive models for different 
  membrane time constants.  The resonance frequencies of 
  \b{$|\kappa_{\lambda}|$} as a function of the membrane time constant
  \b{$\tau_m$} (see Fig.~\ref{fig10}).}

\label{table2}
\end{table}

\paragraph*{Resistive models with diffusive cytoplasm}

We next considered models where the extracellular medium was resistive
as above, but where the intracellular medium (cytoplasm) was
diffusive, and described by a Warburg impedance.  Figure~\ref{fig7}
shows the spatial and frequency behavior of this model. We can see
that the open-circuit (FO) model shows less attenuation with distance
compared to the closed-circuit (FC) model.  Note that these two models
give opposite variations when the extracellular medium has a zero
resistance: in FC type models, \b{$|F_T|$} attenuates more steeply as
a function of distance when the extracellular impedance increases,
whereas in FO type models, the attenuation becomes less steep.
However, the spatial profile of \b{$|F_T|$} also attenuates less with
a diffusive cytoplasm compared to a resistive cytoplasm.  The latter
result is expected, because the higher the frequency the more the
impedance ``short-cuts'' the membrane in this case.  Note that the
Warburg impedance used in all diffusive models considered here was
applied for frequencies larger than \b{$5 ~Hz $}.

It is interesting to note that in the FC model, a resonance appears
around 24~Hz in the modulus of the transfer function
\b{$\kappa_{\lambda}$} (Fig.~\ref{fig7}A).  In contrast, the FO model
does not display a resonance.

\paragraph*{Resistive cytoplasm with diffusive extracellular medium}

Next, we considered the opposite configuration as previously, namely a
resistive model for the cytoplasm, but a diffusive extracellular
medium. Three sets of parameters were chosen for the extracellular
space.  First, a FO type model with a resistive cytoplasm and a
diffusive extracellular medium described by a Warburg type impedance
(black curve in Fig.~\ref{fig9}), and second, a FC type model with
similar parameters (blue curve in Fig.~\ref{fig8}).  These two models
can be justified if one takes into account the Debye layer at the edge
of the membrane \cite{Deg2010,BedDes2010,BedDes2011a}). The case with
a zero extracellular resistance (short-cut) is also shown for
comparison (red curve in Fig.~\ref{fig9}). The latter model represents
the same limit case for both FO and FC models, and therefore
constitutes the frontier between the two families of curves.

\paragraph*{Fully diffusive cable models}

Next, we have considered the case where both intracellular and
extracellular media are diffusive.  Figure~\ref{fig9} shows the
frequency and spatial profiles of the $V_m$ for such fully diffusive
models.  Taking the FO model with low extracellular impedance
(\b{$|z_e|= r_e$} at \b{$1~Hz$}) leads to large differences with the
FC model (Fig.~\ref{fig9}, black) compared to the FO model (blue) or
the FC model with zero extracellular resistance (red).

We can see that, in FC type models, the larger \b{$z_e$}, the steeper
the transfer function attenuates with distance.  In contrast, in FO
type models, larger \b{$z_e^m$} lead to less attenuation.  This
paradoxical result can be explained as follows: in FC models,
\b{$z_e$} plays as similar role as \b{$z_i$}, such that for large
values of their real part, thermal diffusion attenuates the signal; in
FO models, \b{$z_e^m$} plays a similar role as \b{$r_m$}, and large
values of \b{$|z_e^m|$} limit the leak membrane current, reducing the
attenuation with distance.  Thus, for large \b{$|z_e^m|$}, the
dendrites become more ``democratic'' in the sense that the effect of a
given input will be less dependent on its position on the dendrite.
This is only the case for FO models, however.

As above, the model with zero extracellular resistance represents the
same limit case for both FO and FC models, and therefore constitutes
the frontier between the two models. 

\paragraph*{Resonances with diffusive models}

One interesting finding is that resonances appear in several models
using diffusive extracellular impedances (Figs.~\ref{fig7} and
\ref{fig9}). This type of resonance was studied further in
Fig.~\ref{fig10}, where one can see that a resonance in
\b{$|\kappa_{\lambda}|$} also implies a resonance in \b{$|F_T|$}: The
\b{$V_m$} still attenuates with distance independently of the
frequency, so that we always have \b{$\frac{\partial |V_m|}{\partial
    x}<0$}. In addition, Eq.~\ref{eq23} shows that
\b{$|\frac{\partial^2V_m}{\partial x^2}|=|\kappa_{\lambda}||V_m|$}, so
that the quantity \b{$|\frac{\partial^2V_m}{\partial x^2}|$} increases
when \b{$|\kappa_{\lambda}|$} increases with frequency, which implies
that \b{$\frac{\partial |V_m|}{\partial x}$} becomes more negative
because this derivative is always negative.  It follows that
\b{$|F_T|$} attenuates more steeply with distance when
\b{$|\kappa_{\lambda}|$} increases with frequency.  Using a similar
reasoning, one can show that \b{$|F_T|$} attenuates less steeply with
distance when \b{$|\kappa_{\lambda}|$} diminishes with frequency.  We
conclude that the rate of variation of \b{$|F_T|$} with frequency is
always opposed to that of \b{$|\kappa_{\lambda}|$}. Consequently, the
resonance frequency must be the same for \b{$|\kappa_{\lambda}|$} and
\b{$\frac{d|F_T|}{df} =0 $} because we have \b{$\frac{d|F_T|}{df} \leq
  0 $} when \b{$\frac{d|\kappa_{\lambda}|}{df} \geq 0 $} and
\b{$\frac{d|F_T|}{df} \geq 0 $} when
\b{$\frac{d|\kappa_{\lambda}|}{df} \leq 0 $}.

We also see that the peak frequency of the resonance continuously
depends on the membrane time constant (not shown). For example, for
\b{$\tau_m=5~ms$}, the resonance is at about \b{$24~ Hz$}, and for
\b{$\tau_m=20~ms$}, the resonance is at about \b{$8~ Hz$} (for more
details see Fig.~\ref{fig10} and Table~\ref{table2}).  It is
interesting to note that we have observed resonances only in FC type
models with resistive extracellular media and diffusive cytoplasm (see
Fig.~\ref{fig7}), but resonances are present in the two types of
models (FO and FC) when they are fully diffusive.


\section{Discussion}

In this paper, we have introduced a generalization of cable equations
to membranes within media with complex or heterogeneous electrical
properties.  We have shown that generalized cable equations can treat
a number of problems presently not treatable by the traditional cable
equations.  We have shown that the nature of the extracellular medium
has a significant influence on fundamental neuronal properties, such
as \corr{voltage} attenuation with distance, and the spectral profile
of the transmembrane potential.  We enumerate below the consequences
and predictions of this work, as well as outline directions for future
studies.

\corr{A first main result of this paper is to generalize cable
  equations to describe membranes in complex and heterogeneous media.}
To solve this problem, we have introduced the concept of {\it
  generalized current}, and show that the generalized current is
conserved in all situations.  This stands in contrast with the
free-charge current, which is conserved only in special cases.  For
example, if the medium is electrically non-homogeneous (with
conductive and non-conductive domains), there will be charge
accumulation and non-conservation of the free-charge current.  Thus
the traditional cable formalism, which is based on the free-charge
current, cannot treat this problem.  With the generalized current,
however, this problem can be treated in a physically plausible way, in
accordance with Maxwell equations.

One drawback of generalized cable equations is that they cannot be
solved with available neural simulation environments, such as NEURON
\cite{Hines97}, which implements the traditional cable formalism.
Consequently, we have developed a specific method for the numerical
simulation of generalized cables.  This method is implementable with
traditional simulation programs, such as MATLAB.  Further work would
be needed to determine if generalized cable equations could be
included in neural simulators, as a special case.

Note that specialized models different from the standard model were
introduced relatively recently \cite{Poz,Monai} to include aspects
which cannot be treated by the standard model. In \cite{Poz}, the
cytoplasm was considered as non-resistive but capacitive, and was
modeled by a RC circuit.  It was estimated that this capacitive aspect
is important to understand the nature of thermal noise in thin
dendritic branches.  \cite{Monai} considers the case of the
interaction between closely located dendritic branches.  In this case,
the authors study the phenomenon of surface polarization (see also
\cite{BedDes2006b}) and evaluate the magnitude of the Maxwell-Wagner
time of the effective impedance of the extracellular medium, needed to
have significant influences over the attenuation profile of the V$_m$.
These two studies show that the physical nature of the intracellular
or extracellular media can have significant influences on cable
properties.  However, they do represent very particular cases, which
motivated the present study where we have attempted to consider a
broad range of cases, including both intracellular and extracellular
media, as well as ionic diffusion, which was not treated previously.
Thus, the present study generalizes those prior studies.

A second main result of this paper was to also generalize the
electrical circuit representing neuronal membranes.  Instead of
considering the neuron as a closed system, where all outward currents
return to the neuron, we have considered the more general case which
allows current exchange between neighboring neurons, and thus each is
represented by an {\it open circuit}.  We have systematically compared
open-circuit (FO) models with the traditional closed-circuit (FC)
models, and found some important differences. FO models have a
transfer function that depends much less on frequency and space,
compared to FC models (see Figs.~\ref{fig7} and \ref{fig8}).

We also showed that a previously introduced model of non-ideal
cable~\cite{BedDes2008} is equivalent to a traditional cable with
appropriately scaled extracellular resistances (for frequencies
smaller than \b{$100~Hz$}; see Figs.~7 and 8 in \cite{BedDes2008}, as
well as the discussion in that paper).

One of the most important result \corr{of this paper is the finding
  that} the nature of extracellular or intracellular media can have a
strong impact on cable properties such as \corr{voltage} attenuation
with distance.  We have observed that the nature of the extracellular
medium has an opposite impact on distance attenuation on FO and FC
models.  In FO models, larger extracellular impedances lead to less
attenuation and electrotonically more compact dendrites. The
attenuation can be remarkably diminished for fully resistive FO
models, with only a few percent attenuation (Fig.~\ref{fig9}), whereas
for FC type models, the opposite was seen, the dendrites become more
compact for low extracellular impedances. We can say that in these
cases, the effect of distal inputs is close to that of proximal
inputs, and thus the dendrite is more ``democratic''.  It may be that
this remarkable property is present in some types of neurons to reduce
the attenuation of distal inputs, which constitutes another
interesting direction to explore in future work.

Another interesting observation is that diffusive extracellular
impedances can give rise to resonance frequencies (see
Figs.~\ref{fig7} and \ref{fig9}), which also appears as a resonance in
\b{$|\kappa_{\lambda}|$} (Fig.~\ref{fig10}).  The resonant frequency
depends on the membrane time constant, and is in the range of 5-40~Hz,
which is well within the frequency range of brain oscillations such as
theta, alpha, beta or gamma rhythms~\cite{Buzsaki}.  It is therefore
possible that this resonance plays a role in the \corr{genesis of
  oscillatory activity by single neurons}.

Interestingly, we observed that the input impedance of the
extracellular medium (\b{$z_e^{(m)}$}) must necessarily be negative in
the standard model where the medium is resistive. In a closed-circuit
configuration, this means that one must necessarily assume a
  source of energy, such as an electromotive force. This source of
energy can be simulated by a negative impedance.  This important point
was pointed in previous work, where it was called ``anomalous
impedance'' \cite{Conciauro, Mauro}.  Interestingly, this constraint
disappears in the open-circuit configuration.  If the current field is
open in the extracellular medium, then it is not necessary to assume
that \b{$z_e^{(m)}$} is negative, and there is no need of such a
source of energy.

\corr{Finally, while our analysis shows that the nature of the
  extracellular or intracellular media may be influential on
  single-neuron behavior, we can also foresee consequences at the
  network level.  First, the resonance found for some of the media may
  introduce a bias in the genesis of oscillatory behavior by
  populations of neurons.  The fact that the resonance frequency only
  depends on membrane parameters, but not on structural parameters
  such as cell size, suggests that different neurons in the network
  will have the same resonance frequency.  It is thus conceivable that
  population oscillatory activity may occur at this resonance
  frequency. Second, the fact that the diffusive properties of media
  were found to be particularly impactful on the attenuation of distal
  inputs suggests that any regulation of these properties could have
  drastic consequences at the network level.  If diffusive properties
  are modified, for example by glial cells who are known to regulate
  extracellular ionic concentrations \cite{Walz89,Kettenmann95}, it
  may affect the voltage attenuation of all cells in the network and
  therefore change network behavior.}

In conclusion, we think that the generalized cable equations allow one
to treat the problem of how neuronal membranes behave in complex
extracellular and heterogeneous media.  Given the possible strong
impact of such media as found here, future studies should evaluate in
more depth whether such media are indeed influential.  A possible
approach would be to find ``signatures'' of the extracellular medium
from the power spectral density of experimentally observable
variables, such as the membrane potential (for a related approach, see
\cite{BedDes2010}).  The direct measurement of the extracellular
impedance, at present bound to contradictory experimental results
\cite{Gabriel1996a,Gabriel1996b,Logothetis}, should give a definite
indication whether the generalized cable is a necessary approach to
accurately model neurons.

\clearpage
\begin{appendix}

\section*{Appendices}

\corr{\section{Generalized current and charge conservation}}
\label{appgencurr}

\corr{In this appendix, we derive the charge conservation laws for
  different definitions of currents (see Eqs.~\ref{eq1} and
  \ref{eq3}).  Consider a domain \b{$\mathcal{D}$} delimited by a
  closed surface \b{$\partial\mathcal{D}$}.  If we assume that the
  medium and the field are sufficiently regular, then the divergence
  theorem applies in \b{$\mathcal{D}$}, and we have:} \b{
 \begin{equation}
\oiint\limits_{\partial \mathcal{D}}
 \nabla\times\vec{H}\cdot\hat{n}~dS
 \equiv\iiint\limits_{\mathcal{D}}
\nabla\cdot(\nabla\times\vec{H})~ dv \equiv 0
\label{eqA1}
 \end{equation}}
\corr{because the following equality always applies:
\b{$\nabla\cdot(\nabla\times\vec{H})\equiv 0$}~\footnote{We use the
symbol \b{$\equiv$} for a mathematical identity while the symbol
\b{$=$} will mark an equality or a physical law.}}

\corr{From Eqs.~\ref{eq1}, \ref{eq3}  and \ref{eqA1}, we have the following 
identity:}
\b{
\begin{equation}
 \oiint\limits_{\partial \mathcal{D}}\vec{j}^{~g}\cdot\hat{n}~dS \equiv 
 \iiint\limits_{\mathcal{D}}\nabla\cdot\vec{j}^{~g}~dv
=0 ~ ,
\label{eq4}
\end{equation}}
\corr{which is valid for an arbitrary domain \b{$\mathcal{D}$}.}

\corr{One can distinguish three different types of current, the
  generalized current \b{$i^{~g}$}, the current due to free charges
  \b{$i^{~f}$}, and the displacement current \b{$i^{~d}$}. These
  currents can be defined across an arbitrary surface
  \b{$\mathcal{S}$}, according to:} \b{
\begin{equation}
\left \{
\begin{array}{cccc}
 i^{~g} & \overset{\text{def}}{=} & \iint\limits_{ \mathcal{S}}\vec{j}^{~g}\cdot\hat{n}~dS \\\\
 i^{~f} & \overset{\text{def}}{=} & \iint\limits_{ \mathcal{S}}\vec{j}^{~f}\cdot\hat{n}~dS \\\\
 i^{~d} & \overset{\text{def}}{=} &\iint\limits_{ \mathcal{S}}\frac{\partial \vec{D}}{\partial t}\cdot\hat{n}~dS 
\end{array}
\right .
\label{eq5}
\end{equation}}

\corr{Within these definitions, we can write that the generalized
  current \b{$i^{~g}$} is conserved at every time and independently of
  the nature of the medium.  At every time, the inward current
  entering a given domain \b{$\mathcal{D}$} is always equal to the
  outward current exiting that domain, independently of the
  homogeneous or heterogeneous nature of the medium.  It is also
  independent of the fact that there may be charge accumulation in
  some elements of volume, because Eq.~\ref{eq4} always applies.}

\corr{Note that this generalized current conservation law does not
  express anything new on a physical point of view, but is the charge
  conservation law expressed as a function of currents.  Indeed,
  taking into account Maxwell-Gauss law (\b{$\nabla\cdot\vec{D} =
    \rho^{~f}$}), the definition of \b{$\vec{j}^{~g}$} (Eq.~\ref{eq3})
  and the identity given by Eq.~\ref{eq4}, we obtain the differential
  charge conservation law:}  \b{
\begin{equation}
 \nabla\cdot\vec{j}^{~g} =  \nabla\cdot\vec{j}^{~f} + 
\nabla\cdot\frac{\partial \vec{D}}{\partial t}=
\nabla\cdot\vec{j}^{~f} +
\frac{\partial ~\nabla \cdot\vec{D}}{\partial t}=
\nabla\cdot\vec{j}^{~f} + \frac{\partial \rho }{\partial t}^{f}=0
\label{eq6}
\end{equation}}

\section{Displacement current, free current and charge accumulation}
\label{appA}

In this appendix, we show explicitly that the displacement current
\b{$i^{~d}$} can be used to formally calculate the charge variation
in a given domain \b{$\mathcal{D}$}. 
Moreover, we show that the displacement current across a closed
surface \b{$\partial \mathcal{S}$} which surrounds a given domain
\b{$\mathcal{D}$} is zero when there is no charge variation inside
the domain.  

By definition, the density of displacement current (Eq.~\ref{eq5})
in frequency space is given by:
\b{
\begin{equation}
 \vec{j}^{~d}(\vec{x},\omega ) =  i\omega\varepsilon (\vec{x},\omega )\vec{E}(\vec{x},\omega )
\label{eqa1}
\end{equation}}
where \b{$\omega = 2\pi f$}.
By applying the divergence on \b{$\vec{j}^{~d}$} and taking into
account Maxwell-Gauss law, we obtain:
\b{
\begin{equation}
\nabla\cdot \vec{j}^{~d} =  i\omega\nabla\cdot(\varepsilon \vec{E})=
i\omega\rho^{f}
\label{eqa2}
\end{equation}}
Thus, we can calculate
the amount of free charges in a given domain \b{$\mathcal{D}$} from the
density of displacement current in frequency space.  To do this,
we have
\b{
\begin{equation}
 Q^{f}(\omega)=\iiint\limits_{\mathcal{D}}\rho^{f}(\vec{x},\omega)~dv
= \frac{1}{i\omega}\iiint\limits_{\mathcal{D}}\nabla\cdot\vec{j}^{~d}(\vec{x},\omega)~dv
\equiv
\frac{1}{i\omega}\oiint\limits_{\partial\mathcal{D}}\vec{j}^{~d}\cdot\hat{n}~dS
= \frac{i^{~d}(\omega)}{i\omega}
\label{eqa3}
\end{equation}}
where \b{$i^{~d}$} is the displacement current flowing across
surface \b{$\partial \mathcal{S}$}.  Applying the inverse Fourier transform,
we obtain the rate of free charge variation in domain \b{$\mathcal{D}$}:
\b{
\begin{equation}
 \frac{dQ}{dt}^{f}(t)=
 i^{~d}(t)
 \label{eqa4}
\end{equation}}
Therefore, one can say that the charge in the considered volume
does not vary \corr{if} the displacement current across surface 
\b{$\partial \mathcal{D}$} is zero.  Finally, because the 
differential conservation law for free charges implies:
\b{
\begin{equation}
\frac{dQ}{dt}^{f}(t)= \iiint\limits_{\mathcal{D}}\frac{\partial \rho^f (\vec{x},t)}{\partial t}~dv =-\iiint\limits_{\mathcal{D}}\nabla\cdot\vec{j}^{~f}(\vec{x},t)~dv
\equiv
-\oiint\limits_{\partial\mathcal{D}}\vec{j}^{~f}\cdot\hat{n}~dS
= -i^{~f}(t) ~ ,
 \label{eqa6}
\end{equation}}
we can then write:
\b{
\begin{equation}
 i^{~g}(t) = i^{~d}(t)+i^{~f}(t)=0
  \label{eqa7}
\end{equation}}
when the surface is closed and when the free charge conservation
law applies.

Thus, the generalized current entering a given closed surface
\b{$\partial\mathcal{D}$} is always equal at every time to the
generalized current exiting \b{$\partial\mathcal{D}$}, {\it even if
there is free charge accumulation} inside
\b{$\partial\mathcal{D}$}.  However, this equality does not allow
one to deduce if there are variations of free charge density inside
\b{$\partial\mathcal{D}$}, because the displacement current must
necessary be zero across \b{$\partial\mathcal{D}$} to have
\b{$\frac{dQ}{dt}^f=0$} (see Eq.~\ref{eqa4}).  In other words, it
is necessary that the displacement current entering
\b{$\partial\mathcal{D}$} is equal to the displacement current
exiting \b{$\partial\mathcal{D}$} to have a constant charge inside
\b{$\partial\mathcal{D}$}.  Note that in any given circuit,
Kirchhoff's current law always applies to the generalized current,
even if there is charge accumulation inside the circuit, whereas it
applies to the free charge current {\it only assuming there is no
charge accumulation inside the circuit}.

\section{Input impedance of a cable segment in series with an
arbitrarily complex impedance} 
\label{appB}

In this appendix, we calculate the input impedance of a cable
segment of length \b{$l$} when this segment is connected to an 
arbitrary impedance \b{$Z_a$} (see Fig.~\ref{fig4}).

By definition, we have in \b{$x=0$}:
\b{
\begin{equation}
Z_{in}^l[Z_a] = \frac{V_m~(0,\omega)}{i_i^{~g}(0,\omega)}
\label{eqb1}
\end{equation}}

Applying Eq.~\ref{eq25} allows us to directly express \b{$V_m$}
as a function of the cable parameters.  We have
\b{
\begin{equation}
 V_m(0,\omega) = A^{+}(\omega)~e^{\kappa_{\lambda} l}
+A^{-}(\omega)~e^{-\kappa_{\lambda} l},
\label{eqb2}
\end{equation}}
Similarly, applying Eqs.~\ref{eq14}, \ref{eq17} and \ref{eq22}, we obtain:
\b{
$$
i_i^{~g} = -\frac{1}{z_i}[1+\frac{z_e^{(m)}}{r_m}((1+i\omega\tau_m)]\frac{\partial V_m}{\partial x}
$$
}
This last expression allows us to express the current
at coordinate \b{$x=0$} as a function of the cable parameters:
\b{
\begin{equation}
 i_i^{~g}(0,\omega) = \frac{\kappa_{\lambda}}{\bar{z}_i}[A^{+}(\omega)~e^{\kappa_{\lambda}l}
-A^{-}(\omega)~e^{-\kappa_{\lambda}l}]
\label{eqb3}
\end{equation}}
where \b{
\begin{equation}
\bar{z}_i = \frac{z_i}{1+\frac{z_e^{(m)}}{r_m}(1+i\omega\tau_m)}
\label{eqb4} 
\end{equation}}
Thus, the expression for the input impedance \b{$Z_{in}^{l}$} is
given by:
\b{
\begin{equation}
Z_{in}^{l}[Z_a] = \frac{\bar{z}_i}{\kappa_{\lambda}}\cdot
\frac{(\frac{A^{+}}{A^{-}})\cdot e^{2\kappa_{\lambda}l}+1}{(\frac{A^{+}}{A^{-}})\cdot e^{2\kappa_{\lambda}l}-1}
\label{eqb5}
\end{equation}}
We can then evaluate the ratio \b{$\frac{A^{+}}{A^{-}}$} by using
the conditions of continuity of the current and of the voltage at
point \b{$x=l$}.  Applying Eqs.~\ref{eq25} and \ref{eq14} to that
point gives:
\b{
\begin{equation}
\left \{
\begin{array}{ccccc}
V_m (l,\omega ) &=& A^{+}(\omega) + A^{-}(\omega)&(a) \\\\
i_i^{~g} (l,\omega ) &=& \frac{\kappa_{\lambda}}{\bar{z}_i}~[A^{+}(\omega) - A^{-}(\omega)]&(b)
\end{array}
\right .
\label{eqb6}
\end{equation}}
Thus, we have 
\b{
\begin{equation}
Z_a = \frac{V_m (l,\omega )}{i_i^{~g} (l,\omega )}
= \frac{\frac{A^{+}}{A^{-}}+1}{\frac{\kappa_{\lambda}}{\bar{z}_i}~[\frac{A^{+}}{A^{-}}-1]}~,
\label{eqb7}
\end{equation}}
and we can write
\b{
\begin{equation}
\frac{A^{+}}{A^{-}}=\frac{\kappa_{\lambda}Z_a + \bar{z}_i}{\kappa_{\lambda}Z_a - \bar{z}_i}
\label{eqb8}
\end{equation}}
It follows that the input impedance \b{$Z_{in}^l$} is given by:
\b{
\begin{equation}
Z_{in}^l[Z_a] = \frac{\bar{z}_i}{\kappa_{\lambda}}~
\frac{(\kappa_{\lambda}Z_a + \bar{z_i})~e^{2\kappa_{\lambda}l} +( \kappa_{\lambda}Z_a-\bar{z}_i)}{(\kappa_{\lambda}Z_a +\bar{z}_i )~e^{2\kappa_{\lambda}l} -(\kappa_{\lambda}Z_a -\bar{z}_i) }
\label{eqb9}
\end{equation}}
where
\b{
$$
\bar{z}_i = \frac{z_i}{1+\frac{z_e^{(m)}}{r_m}(1+i\omega\tau_m)}
$$
}

Note that \b{$Z_{in}^l[Z_a] \rightarrow
\frac{\bar{z}_i}{\kappa_{\lambda}}$} when \b{$l\rightarrow \infty$}, and
\b{$Z_{in}^l[Z_a] \rightarrow
\frac{\bar{z}_i}{\kappa_{\lambda}}~coth(\kappa_{\lambda}l)$} when
\b{$Z_a\rightarrow \infty$}.

\section{Calculation of the transfer function $F_T$}
\label{appC}

In this appendix, we calculate the transfer function
\b{$F_T(l,\omega;Z_a)=\frac{V_m(l,\omega )}{V_m(0,\omega )}$}
using the same conditions and conventions as for Appendix~\ref{appB}.

Applying Eq.~\ref{eqb5}a gives: 
\b{
\begin{equation}
\left \{
\begin{array}{ccccc}
V_m (0,\omega ) &=& A^{+}(\omega )~
e^{\kappa_{\lambda}l} &+& A^{-}(\omega )~
e^{-\kappa_{\lambda}l}\\\\
V_m (l,\omega ) &=& A^{+}(\omega ) &+& A^{-}(\omega )
\label{eqc1}
\end{array}
\right .
\end{equation}}
Thus, we have
\b{
\begin{equation}
F_T(l,\omega;Z_a) = \frac{A^{+}(\omega ) + A^{-}(\omega )}
{A^{+}(\omega )~
e^{\kappa_{\lambda}l} + A^{-}(\omega )~
e^{-\kappa_{\lambda}l}}
\label{eqc2}
\end{equation}}
Applying Eq.~\ref{eqb7} gives the transfer function:
\b{
\begin{equation}
F_T(l,\omega ; Z_a) = \frac{\kappa_{\lambda}Z_a}{\kappa_{\lambda}Z_a~cosh (\kappa_{\lambda}l) +\bar{z}_i sinh(\kappa_{\lambda}l)}
\label{eqc3}
\end{equation}}
where
\b{
$$
\bar{z}_i = \frac{z_i}{1+\frac{z_e^{(m)}}{r_m}(1+i\omega\tau_m)}
$$
}

Note that \b{$F_T(l,\omega ; 0)=0$} and \b{$F_T(l,\omega
;\infty)=\frac{1}{cosh (\kappa_{\lambda}l)}$}.

\begin{figure}[bh!] 
\centering
\includegraphics[width=10cm]{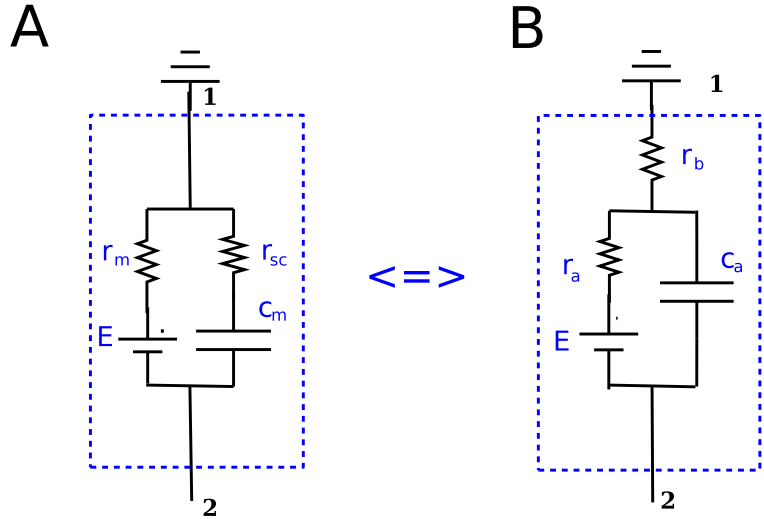}

\caption{\small \corr{(Color online)} Equivalence of the electrical
  circuits of open-circuit and non-ideal cable models.  The circuits A
  and B are equivalent when the ratio \b{$\frac{V_{12}(\omega
      )}{I(\omega)}$} of the voltage difference between points \b{1}
  and \b{2} and the input current between these points is invariant,
  and when the correspondence between the elements of these circuits
  are independent of frequency. Note that the values of the elements
  between the two circuits are related by a transformation law which
  is independent of frequency; this equivalence also applies to the
  temporal domain.  In other words, according to this equivalence, the
  two circuits are equivalent when it is impossible to distinguish
  their topology from external measurements.  The circuit \b{A}
  corresponds to the non-ideal capacitance model introduced previously
  \cite{BedDes2008}, while circuit \b{B} corresponds to a ``standard
  cable model'' with a short-cut (zero extracellular resistivity).}

\label{figF1}
\end{figure}

\section{A new interpretation of the non-ideal cable}
\label{appF}

In this appendix, we show that the non-ideal capacitance model
introduced previously \cite{BedDes2008} is equivalent to an
open-circuit resistive model if we assume that the
circuits \b{A} and \b{B} in Fig.~\ref{figF1} are linked by the
following transformation: \b{
\begin{equation}
\left \{
\begin{array}{cccccccc}
r_a &=& r_m- \frac{r_mr_{sc}}{r_m+r_{sc}}&~~~~ & r_m &=& r_a+r_b  \\\\
r_b &=& \frac{r_mr_{sc}}{r_m+r_{sc}}&~~~~ & r_{sc} &=& r_b+ \frac{r_b^2}{r_a} \\\\
c_a &=& \frac{(r_m+r_{sc})}{r_m-\frac{r_mr_{sc}}{r_m+r_{sc}}}c_m
&~~~~ & c_m &=& \frac{r_a^2}{(r_a+r_b)^2}c_a 
 \end{array}
\right .
\label{eqf1}
\end{equation}}
We show that the \b{$V_m$} in the non-ideal cable model corresponds
to \b{$V_{i}$} in an open-circuit (FO) type resistive model, 
with a reference located far-away.
According to circuits \b{A} and \b{B} in Fig.~\ref{figF1}, we have:
\b{
\begin{equation}
\left \{
\begin{array}{ccc}
circuit~ A & (r_{sc} \oplus c_m)\parallel r_m \\
circuit~ B & (r_a \parallel c_a)\oplus r_b
 \end{array}
\right .
\label{eqf2}
\end{equation}}
It follows that the impedances of circuits A and B are equal if
we have:  
\b{
\begin{equation}
 \frac{V_{12}(\omega)}{i(\omega)} = \frac{r_m +i\omega r_mr_{sc}c_m}{1+i\omega c_m(r_m+r_{sc})}
=\frac{r_a+r_b +i\omega c_ar_ar_b}{1+i\omega c_ar_a}
\label{eqf3}
\end{equation}} 

We see that the ratio \b{$\frac{V_{12}(\omega)}{i(\omega)} $} is a
homographic transform of variable \b{$\omega$}. Consequently,
\b{$\forall \omega$} we have the following relation
\b{$\frac{V_{12}(\omega)}{i(\omega)}
  =\frac{a_A+b_A\omega}{1+d_A\omega}=\frac{a_B+b_B\omega}{1+d_B\omega
  }$} when the two circuits are equivalent.  The only way to guarantee
that the equivalence is independent of frequency is to assume that the
corresponding coefficient of the transformations are equal.  We can
thus set \b{$a_A = a_B$}, \b{$b_A = b_B$} and \b{$d_A = d_B$}.  This
gives us 3 equations which link the 3 parameters of circuit A to those
of circuit B.  The solution is the transformation law
(Eqs.~\ref{eqf1}).  Thus, on a physical point of view, one cannot
distinguish the topology of circuits A and B if we would perform
external measurements.  Moreover, because the functions \b{$r_m =
  f_m(r_a,r_b.r_c)$}, \b{$r_{sc} = f_{sc}(r_a,r_b.r_c)$}, \b{$r_{c_m}
  = f_{c_m}(r_a,r_b.r_c)$} do not depend on frequency, their
equivalence will be also valid for all frequencies.  We can deduce
that the two circuits will behave identically as a function of time
.

It follows that the \b{$V_m$} (between points 1 and 2) in circuit A
(non-ideal capacitance) corresponds to the \b{$V_{i}$} relative to a
far-away reference in circuit B (see Table~\ref{table1}).  Therefore,
a model with non-ideal capacitance and zero extracellular resistance
should produce a \b{$V_{m}$} equivalent to the \b{$V_{i}$} of a model
with ideal capacitance and resistive extracellular medium. Thus, the
frequency-scaling behavior of the \b{$V_{m}$} obtained in a previous
non-ideal cable model \cite{BedDes2008} also applies to the resistive
FO model, but only if one studies the intracellular potential
\b{$V_{i}$}.

\end{appendix}


\subsection*{Acknowledgments}

Research supported by the CNRS, the ANR (Complex-V1 project)
  and the European Union (BrainScales FP7-269921 and Magnetrodes
  FP7-600730).


\label{Bibliographie-Debut}


\begin{thebibliography}{99}

\small

\bibitem{Rall1962} Rall, W. (1962) Electrophysiology of a dendritic
neuron model.  {\it Biophys J.} {\bf 2}: 145-167.

\bibitem{Rall1995} Rall, W. (1995) \textit{The theoretical
foundations of dendritic function}. MIT Press, Cambridge, MA.

\bibitem{Logothetis} Logothetis N.K., Kayser, C. and Oeltermann, A.
(2007) In vivo measurement of cortical impedance spectrum in
monkeys : implications for signal propagation. {\it Neuron} {\bf
55}: 809-823.

\bibitem{Gabriel1996a} Gabriel, S., Lau, R.W. and Gabriel, C.
(1996a) The dielectric properties of biological tissues : I.
Literature survey. {\it Phys. Med. Biol.}. {\bf 41 }: 2231-2249.

\bibitem{Gabriel1996b} Gabriel, S., Lau, R.W. and Gabriel, C.
(1996b) The dielectric properties of biological tissues : II.
Measurements in the frequency range 10 Hz to 20 GHz. {\it Phys. Med.
Biol.}. {\bf 41}: 2251-2269.

\bibitem{BedDes2006a} B\'edard, C., H. Kr\"oger, and A. Destexhe.
  2006.  Does the 1/f frequency scaling of brain signals reflect
  self-organized critical states~? {\it Physical Review Lett.} {\bf
    97}: 118102.

\bibitem{Baz2011} Bazhenov M, Lonjers P, Skorheim P, Bedard C and
  Destexhe A.  (2011) Non-homogeneous extracellular resistivity
  affects the current-source density profiles of up-down state
  oscillations {\it Phil Trans R Soc A} {\bf 369}: 3802-3819.

\bibitem{BedDes2010} B\'edard, C., Rodrigues, S., Roy, N.,
Contreras,  D. and  Destexhe, A. (2010) Evidence for
frequency-dependent extracellular impedance from the transfer
function between extracellular and intracellular potentials.  {\it
J. Computational Neurosci.} {\bf 29}: 389-403.

\bibitem{Deg2010} Dehghani, N., B\'edard, C., Cash, S.S., Halgren,
E. and Destexhe, A. (2010) Comparative power spectral analysis of
simultaneous elecroencephalographic and magnetoencephalographic
recordings in humans suggests non-resistive extracellular media. 
{\it J. Computational Neurosci.} {\bf 29}: 405-421.

\bibitem{BedDes2011a} B\'edard, C. and Destexhe, A. (2011) A
generalized theory for current-source density analysis in brain
tissue. {\it Physical Review E} {\bf 84}: 041909.

\bibitem{BedDes2006b} B\'edard, C., Kr\"oger, H. and Destexhe, A.,
(2006b) Model of low-pass filtering of local field potentials in
brain tissue. {\it Phys. Rev. E}~{\bf 73}:\mbox{051911}.

\bibitem{BedDes2009}B\'edard, C. and Destexhe, A. (2009)
Macroscopic models of local field potentials and the apparent 1/f
noise in brain activity. {\it Biophys. J.} {\bf 96}: 2589-2603.

\bibitem{Hines97} Hines, M.L. and Carnevale,  N.T (1997) The NEURON
simulation environment. {\it Neural Computation} {\bf 9}: 1179-1209.

\bibitem{Tuckwell} Tuckwell, H.C. (1988) \textit{Introduction to
Theoretical Neurobiology: Linear cable theory and dendritic
structure}. Cambridge University Press, Cambdridge, UK.

\bibitem{BedDes2008}B\'edard, C. and Destexhe, A. (2008) A modified
cable formalism for modeling neuronal membranes at high
frequencies.  {\it Biophys. J.} {\bf 94}: 1133-1143.

\bibitem{Wu} Johnston, D. and Wu, S.M. (1995) \textit{Foundations
of cellular neurophysiology}. MIT Press, Cambridge, MA.

\bibitem{Koch} Koch, C. (1999) {\it Biophysics of Computation}. 
Oxford University press, Oxford, UK.

\bibitem{DesBed2012} Destexhe, A. and B\'edard, C. (2012) Do
neurons generate monopolar current sources~? {\it J. 
Neurophysiol.} {\bf 108}: 953-955.

\bibitem{Riera} Riera, J.J., Ogawa, T., Goto, T., Sumiyoshi, A.,
Nonaka, H., Evans, A., Miyakawa, H. and Kawashima, R. (2012)
Pitfalls in the dipolar model for the neocortical EEG sources. 
{\it J. Neurophysiol} {\bf 108}: 956-975. 

\bibitem{BedDes2004} B\'edard, C., Kr\"oger, H. and Destexhe, A. 
(2004) Modeling extracellular field potentials and the
frequency-filtering properties of extracellular space.  {\it
Biophys. J.} {\bf 64}: 1829-1842.

\bibitem{Rieracable} Wang K, Riera JJ, Enjieu-Kadji H and Kawashima R.
  (2013) The role of the extracellular conductivity profiles in
  compartmental models for neurons: Particulars for layer 5 pyramidal
  cells.  {\it Neural Computation} {\bf 25}: 1807-1852.

\bibitem{Poz} Poznanski, R. (2010) Thermal noise due to
surface-charge effects within the Debye layer of endogenous
structures in dendrites. {\it Physical Review E} {\bf 81}: 021902.

\bibitem{Conciauro} Conciauro, G. and Puglisi, M. (1981) Meaning of
the negative impedance. {\it NASA STI/Recon Technical Report} {\bf
82}: 14458.

\bibitem{Mauro} Mauro, A. (1961) Anomalous impedance, a
phenomenological property of time-variant resistance: An analytic
review. {\it Biophys. J.} {\bf 1}: 353-372.

\bibitem{Braitenberg} Braitenberg, V. and A. Shutz. (1998) {\it
Cortex: Statistics and Geometry of Neuronal Connectivity.} (2nd ed.),
Springer-Verlag, Berlin, Germany.

\bibitem{Nicho2005} Nicholson, C.. (2005)  Factors governing
diffusing molecular signals in brain extracellular space.  {\it J.
Neural Transm. } {\bf 112}: 29-44. 

\bibitem{Monai} Hiromu, M, Inoue, M., Miyakawa, H. and Aonishi, T.
  (2012) Low-Frequency dielectric dispersion of brain tissue due to
  electrically long neurites. {\it Physical Review E} {\bf 86}:
  061911.

\bibitem{Buzsaki} Buzsaki G. (2006) {\it Rhythms of the Brain.},
 Oxford University Press, Oxford UK.
  

\bibitem{Walz89} \corr{Walz W. (1989) Role of glial cells in the
    regulation of the brain ion microenvironment.  {\it Progress
      Neurobiol.} {\bf 33}: 309-333.}

\bibitem{Kettenmann95} \corr{Kettenmann H. and Ransom BR. (1995) {\it
      Neuroglia}, Oxford University Press, Oxford, UK.}


\bibitem{George1999} George, M., Lisanby, S., and Sackeim, H. (1999)
  Transcranial magnetic stimulation : Applications in neuropsychiatry.
  {\it Arch. Gen. Psychiatry}, {\bf 56} (4) :300–311.

\bibitem{Landau} Landau, L.D. and ifshitz, E.M. (1984)  {\it
Electrodynamics of Continuous Media.} Pergamon Press, Moscow,
Russia.

\bibitem{Purcell} Purcell, E.M. (1985) {\it Electricity and
Magnetism}.  Berkeley Physics Course Vol.2 chap. 3.

\end{thebibliography}
\end{document}